\documentclass[useAMS,usenatbib]{mn2e}

\title[Multi-epoch pc-scale observations of the blazar
  PKS\,1510-089]{Multi-epoch parsec-scale observations of the blazar PKS\,1510-089}
\author[M. Orienti et al. ]
  {M. Orienti$^{1,2}$\thanks{E-mail: orienti@ira.inaf.it},
T. Venturi$^{2}$, D. Dallacasa$^{1,2}$, F. D'Ammando$^{3}$, M. Giroletti$^{2}$,
\newauthor G. Giovannini$^{1,2}$, S. Vercellone$^{3}$, M. Tavani$^{4,5}$\\
$^1$Dipartimento di Astronomia, Universit\`a di Bologna, via Ranzani 1,
I-40127, Bologna, Italy \\
$^2$INAF -- Istituto di Radioastronomia, via Gobetti 101, I-40129, Bologna,
Italy \\
$^{3}$INAF -- Istituto di Astrofisica Spaziale e Fisica Cosmica, via U. La
Malfa 153, I-90146 Palermo, Italy \\
$^{4}$Dipartimento di Fisica, Universit\`a di Roma ``Tor Vergata'', Via
della Ricerca Scientifica 1, I-00133 Roma, Italy\\
$^{5}$INAF -- Istituto di Astrofisica Spaziale e Fisica Cosmica, 
via Fosso del Cavaliere 100, I-00133 Roma, Italy
}
\date{Received \today; accepted ?}

\pagerange{\pageref{firstpage}--\pageref{lastpage}} \pubyear{2002}

\def\LaTeX{L\kern-.36em\raise.3ex\hbox{a}\kern-.15em
    T\kern-.1667em\lower.7ex\hbox{E}\kern-.125emX}

\begin{document}

\label{firstpage}

\maketitle

\begin{abstract}
We investigate the flux density variability and changes in the
parsec-scale radio 
structure of the flat spectrum radio quasar PKS\,1510-089. This source
was target of multi-epoch Very Long Baseline Interferometer (VLBI) and
Space-VLBI observations at 4.8, 8.4 and 22 GHz carried out between
1999 and 2001. The comparison of the parsec-scale structure observed
at different epochs shows the presence of a non-stationary jet feature
moving with a superluminal apparent velocity of 16.2$c$
$\pm$0.7$c$. Over three epochs at 8.4 GHz during this period 
the core flux density varies of
about 50\%, while the scatter in the jet flux density is within
10\%. The polarization percentage of both core and jet components
significantly change from 2 to 9 per cent, while the polarization angle
of the core shows an abrupt change of about 90 degrees becoming
roughly perpendicular to the jet direction, consistent with a change in the
opacity. To complete the picture of the physical processes at
work, we complemented our observations with multi-epoch Very Long
Baseline Array (VLBA) data at 15 GHz from the MOJAVE programme
spanning a time baseline from 1995 to 2010. Since 1995 jet components
are ejected roughly once per year with the same position angle and an
apparent speed between 15$c$ and 20$c$, 
indicating that no jet precession is taking place on a timescale
longer than a decade in our frame. 
The variability of the total intensity flux
density together with variations in the polarization properties may be
explained assuming either a change between the optically-thick and -thin
regimes produced by a shock that varies the opacity, or a highly
ordered magnetic field produced by the compression of the relativistic
plasma by a shock
propagating along the jet. 
Taking into account the high $\gamma$-ray emission observed from this
source by the AGILE and {\it Fermi} satellites we investigated the 
connection between the radio and $\gamma$-ray activity during
2007-2010. Multi-wavelength flux and polarization observations
suggest that during some $\gamma$-ray flaring episodes the emission
at high ($\gamma$-ray) and low (radio) energies has origin in the same region.
\end{abstract}


\section{Introduction}

Powerful radio emission is rare in the population of 
active galactic
nuclei (AGN). This phenomenon, only found in about 10\% of the AGNs, is 
associated with the presence of relativistic particles produced in the
central regions of the AGN, and then channelled through the jets out to
distances of hundreds of kpc up to a few Mpc. Such relativistic
particles are responsible for the synchrotron radiation at the
origin of the radio emission. Furthermore, they may scatter
low energy photons to high energy bands by inverse Compton (IC)
processes, suggesting a connection between radio and high energy
$\gamma$-ray emission. Indeed, all the $\gamma$-ray AGN detected by EGRET 
are radio-loud sources, strongly supporting this
scenario. No radio-quiet AGNs were firmly detected in $\gamma$-rays
by the Large Area Telescope (LAT) on board the {\it Fermi}
satellite during the first year of operation, although possible 
association for the LAT sources with radio-quiet AGNs are tentatively 
proposed in the first LAT AGN Catalog \citep{abdo10a}. 
As {\it Fermi}-LAT continues to collect data and its sensitivity 
increases, some of these associations could be confirmed. \\
The AGN dominating the $\gamma$-ray sky is 
the blazar population, comprising flat spectrum radio quasars (FSRQ)
and BL Lac objects. These sources are characterized by the presence
of a compact radio core, apparent superluminal jet speed, 
extreme flux density variability in all
bands, and high fraction of polarized optical and radio
emission. Their observational properties are interpreted as the result of
severe beaming effects due to the orientation of the relativistic jet at very
small angles to the line of sight. \\
Thanks to the episodes of enhanced luminosity across
the entire electromagnetic spectrum, it is possible to set constraints
on the physical properties of the region along the jet
responsible for the emission at the various wavelengths. 
Radio monitoring of EGRET sources suggested that the highest levels of
$\gamma$-ray emission is observed close in time to radio flares
\citep[e.g.][]{lathe03}, and connected with the emission of a new
jet component \citep{jorstad01}. Both pieces of evidence give 
support to the idea that the strongest 
$\gamma$-ray emission is strictly related
to a shock in the jet that produces the synchrotron radio flare
\citep[e.g.][]{marscher85}. 
It must be noted that the EGRET data were sparse, 
with large uncertainties and selection effects, precluding a clear  
temporal correlation between
$\gamma$-ray activity and the emission at lower frequencies. 
The advent of the AGILE and {\it Fermi} $\gamma$-ray satellites allowed us 
to test the results obtained in the EGRET era, 
providing details on the connection between $\gamma$-ray and radio
emission to find out the mechanisms responsible for the high-energy emission.
Correlation studies of the first three months of {\it Fermi}-LAT data and 
quasi-simultaneous Very Long Baseline Interferometry (VLBI)
observations showed that the gamma-ray emitting blazars have faster
apparent jet speeds \citep{lister09c},
wider apparent opening angles \citep{pushkarev09}, 
and higher variability and Doppler factor \citep{savolainen10}, with
respect to blazars with weak $\gamma$-ray emission.
Variability studies from radio to high energy bands give important
insight to locate the site of flares and infer the physical
conditions of the jet. In particular, using the most recent $\gamma$-ray 
as well as radio polarimetric data, a likely close connection between 
high $\gamma$-ray states and the activity in the pc-scale core region has
been investigated \citep[see e.g.][]{agudo11}.  In addition, 
Pushkarev et al. (2010) found for a sample of 186 sources a non-zero 
time delay between radio emission measured by pc-scale observations 
at 15 GHz and $\gamma$-ray radiation detected by {\it Fermi}-LAT, 
suggesting that the delay is most likely connected with synchrotron
opacity in the core region. \\
Among blazars, 
PKS\,1510-089 is an ideal target to investigate the location of the
emitting region and the physical processes occurring in 
relativistic jets. This object is a flat spectrum radio quasar at
$z=0.361$ \citep{thompson90} with highly polarized optical emission. 
The synchrotron emission has its peak in the IR band, while the
inverse Compton component peaks in the $\gamma$-ray regime.  
High flux density variability is observed throughout the
electromagnetic spectrum,
from the radio to the
high energy bands. In particular, episodes of high $\gamma$-ray
activity were recently detected by both {\it Fermi}-LAT 
\citep{cutini09,ciprini09,tramacere08}
and the Gamma Ray Imaging Detector (GRID) on board AGILE
\citep{striani10,vercellone09,pucella09,dammando09a,dammando08}. 
Description of
these events, together with their possible connection with
multiwavelengths emission can be found in
\citet{pucella08}, \citet{dammando09b}, \citet{marscher10}, 
\citet{abdo10b}, and \citet{dammando11}.\\   
In the radio band the emission is dominated by the core
component. Multi-epoch parsec-scale observations 
revealed highly superluminal knots with apparent
velocity exceeding 20$c$
\citep[e.g.][]{homan01,jorstad05,lister09b,marscher10}, 
ejected along the north-west direction at an angle of
about -30$^{\circ}$ with respect to the core. 
On arcsecond scale the jet structure is oriented in the opposite
direction, 
indicating a severe misalignment of almost 180$^{\circ}$
between the pc- and kpc-scale jet. Misalignment between pc-
and kpc-scale structure has been frequently observed in core-dominated
sources \citep[e.g.][]{pearson88}. However, misalignment larger
than 110$^{\circ}$
as those found in 0954+556 and 1652+398 \citep{lister01}
is very rare. The extraordinary case of PKS 1510-089 has been
described by \citet{homan02b} assuming a small change of about
12$^{\circ}$-24$^{\circ}$ 
in the intrinsic jet direction, that appears amplified
due to projection effects, providing a simple explanation for the
observed morphology. 
Both the misalignment and the
highly superluminal jet speed indicate that the jet axis of
PKS\,1510-089 forms a very small angle of a few degrees 
to the line of sight. Such an extreme orientation enhances
beaming effects making this source a good target to investigate the
possible connection between $\gamma$-ray flares and the ejection of
jet components. \\
In this paper we present new results of proprietary multi-epoch polarimetric 
VLBI and Space-VLBI observations of PKS\,1510-089
carried out at 4.8, 8.4 and 22 GHz 
between 1999 and 2001 and not yet published, with the aim of 
studying changes in the source structure and investigating
their possible connection with other physical properties, such as 
flux density variability, spectral index distribution and
polarization properties. We then compare our results 
with multi-epoch Very long Baseline Array (VLBA) data at 15 GHz 
from the MOJAVE (Monitoring Of Jets in Active galactic nuclei with
VLBA Experiments) programme\footnote{All the 15-GHz data presented
  in this paper are from the MOJAVE programme. The MOJAVE data archive is 
maintained at http://www.physics.purdue.edu/MOJAVE.} \citep{lister09a}
spanning a larger time interval, 
between 1995 and 2010. The addition of high energy information from AGILE and
{\it Fermi} data gives us a clue to 
connect $\gamma$-ray
emission and radio properties, like flux density variability and 
changes in the pc-scale radio structure for the period 2007--2010.\\
    
\begin{figure}
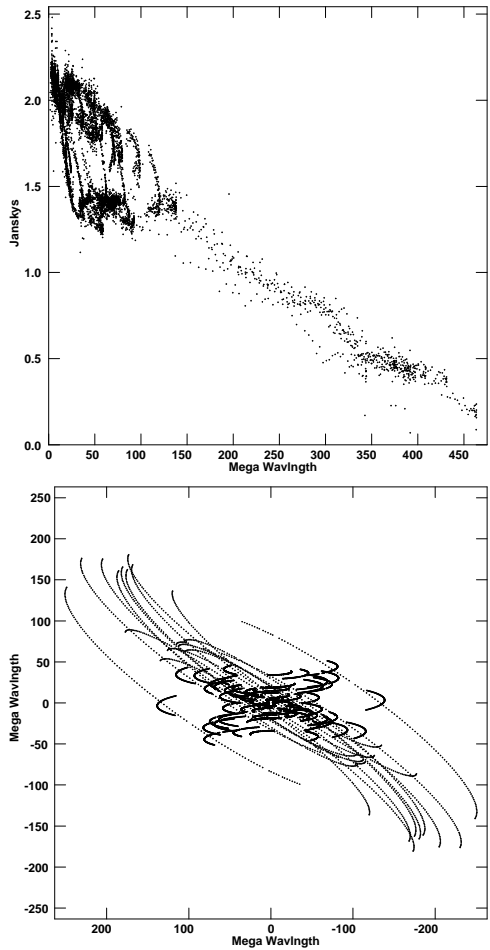

\begin{center}
\includegraphics{1510_BASELINE.PS}
\includegraphics{1510_VSOP_UV.PS}
\vspace{13cm}
\caption{Plot of amplitude vs. projected baseline length at 4.8 GHz
  ({\it top}), and the {\it uv} coverage ({\it bottom}) of Space-VLBI
  (VSOP) observations.}
\label{uv-vsop}
\end{center}
\end{figure}

Throughout this paper, we assume the following cosmology: $H_{0} =
71\; {\rm km \; s^{-1} \; Mpc^{-1}}$, $\Omega_{\rm M} = 0.27$ and
$\Omega_{\rm \Lambda} = 0.73$, in a flat Universe. 
At the redshift of the target 1 arcsec = 5.007
kpc. The spectral index $\alpha$ is defined as $S$($\nu$)
$\propto \nu^{- \alpha }$.\\

\section{Radio data}

\subsection{VLBI and Space-VLBI observations}

PKS\,1510-089 was target of Space-VLBI observations. It was observed at
4.8 GHz (C band) by
VLBA+HALCA on 1999 August 11 and on 2000 May 13 in single polarization
mode. In each observing run the target was observed for about 6 hours. 
Examples of the {\it uv} coverage and amplitude vs baseline length
are presented in Fig. \ref{uv-vsop}.\\
The target source was observed with the VLBA with the
addition of the Effelsberg telescope at 8.4 GHz (X band) and
22.2 GHz (K band), in full polarization mode
with a bandwidth of 32 MHz at 128 Mbps. 
To study possible changes in the source structure, 
subsequent VLBA observations at 8.4 GHz were carried out
in full polarization with a recording
bandwidth of 32 MHz at 128 Mbps. 
The correlation was performed at the VLBA correlator 
in Socorro and the data reduction was carried out with the NRAO AIPS
package. In the X band the instrumental polarization was removed
  by using the AIPS task PCAL. The absolute orientation of the
  electric vector of the calibrator B1749+096 was then compared with
  the VLA/VLBA polarization calibration database to derive the
  corrections. The values derived are in good agreement ($\leq$ 5$^{\circ}$).

\begin{figure*}
\begin{center}
\includegraphics{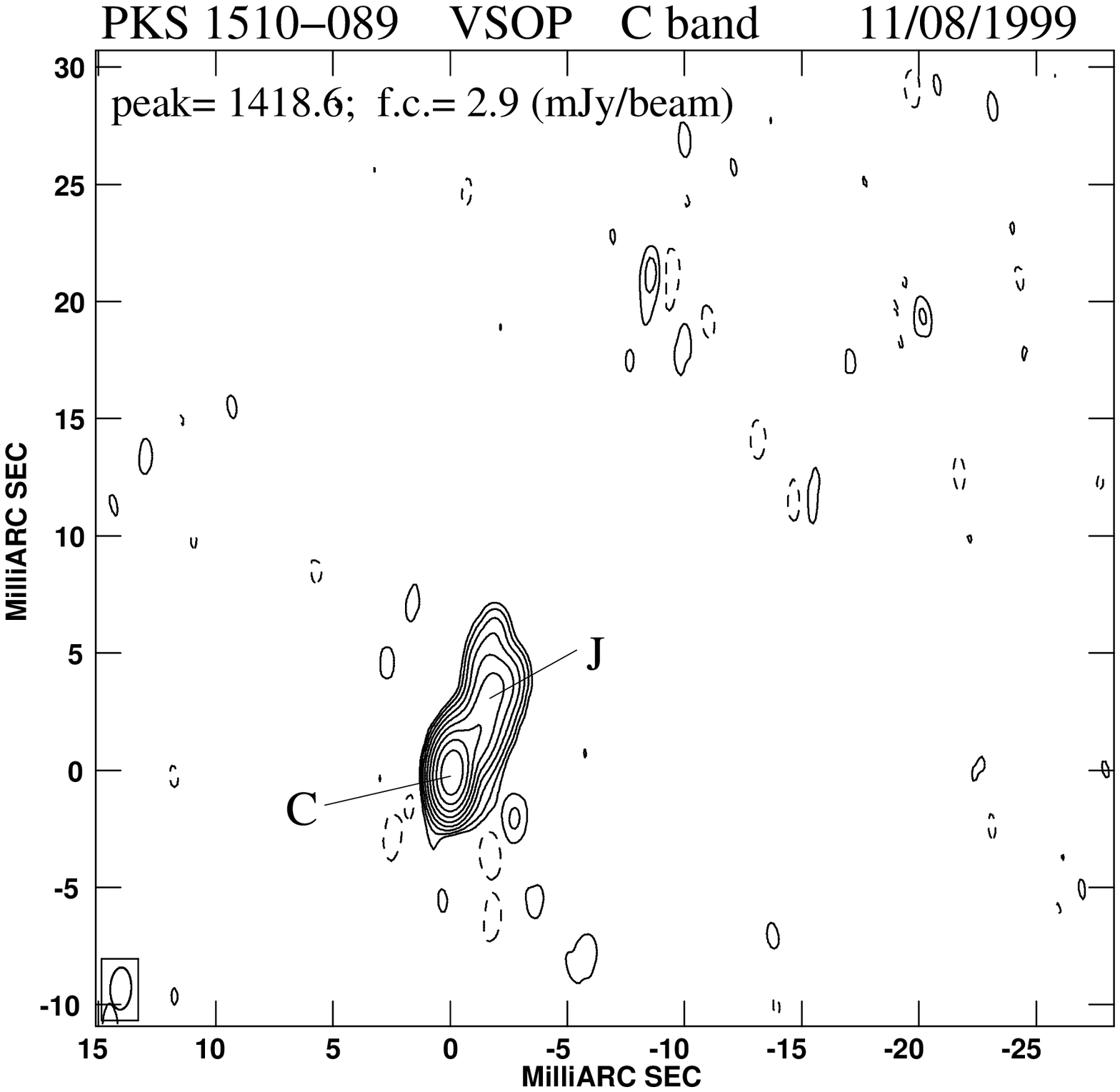}
\includegraphics{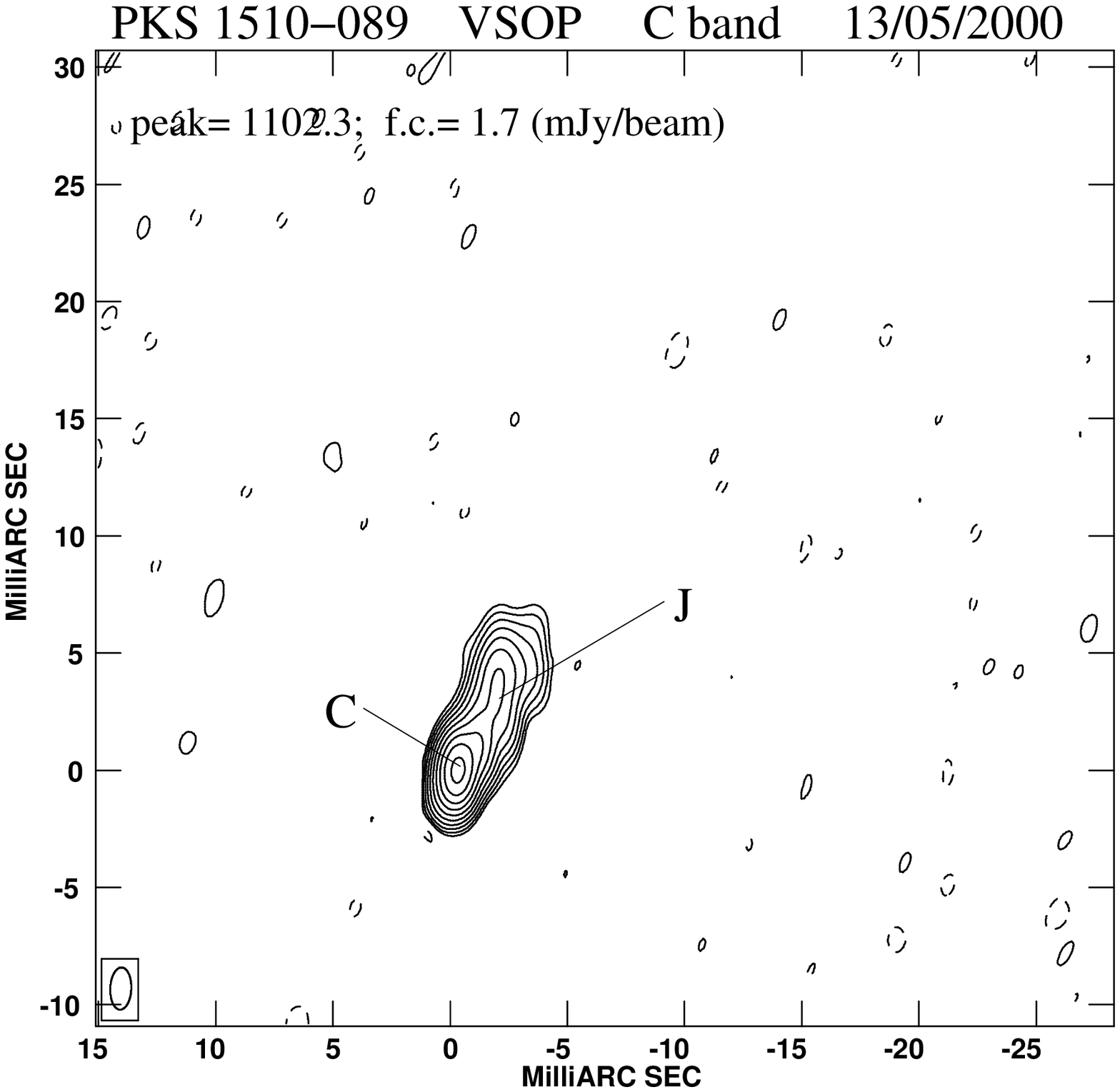}
\includegraphics{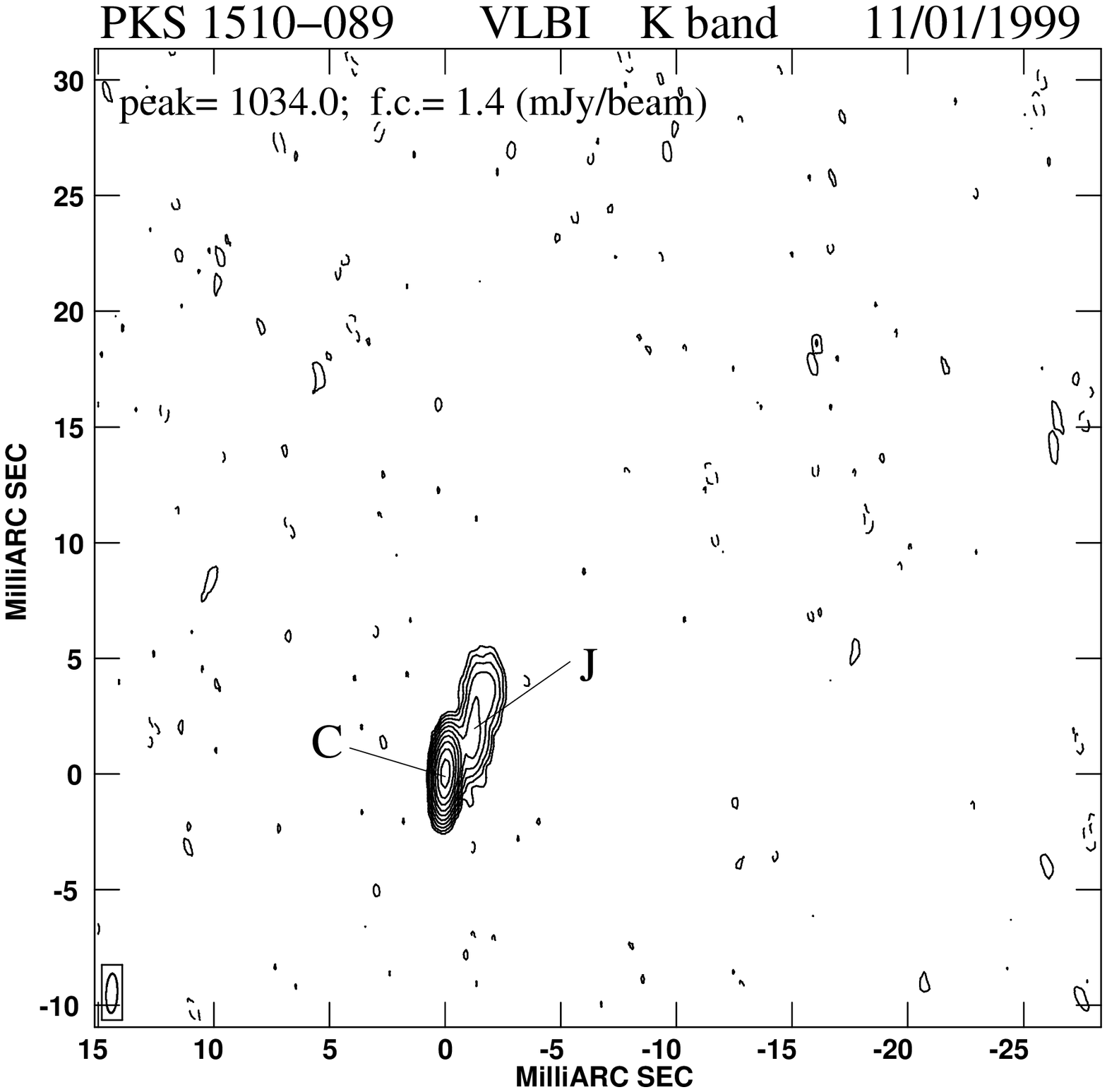}
\includegraphics{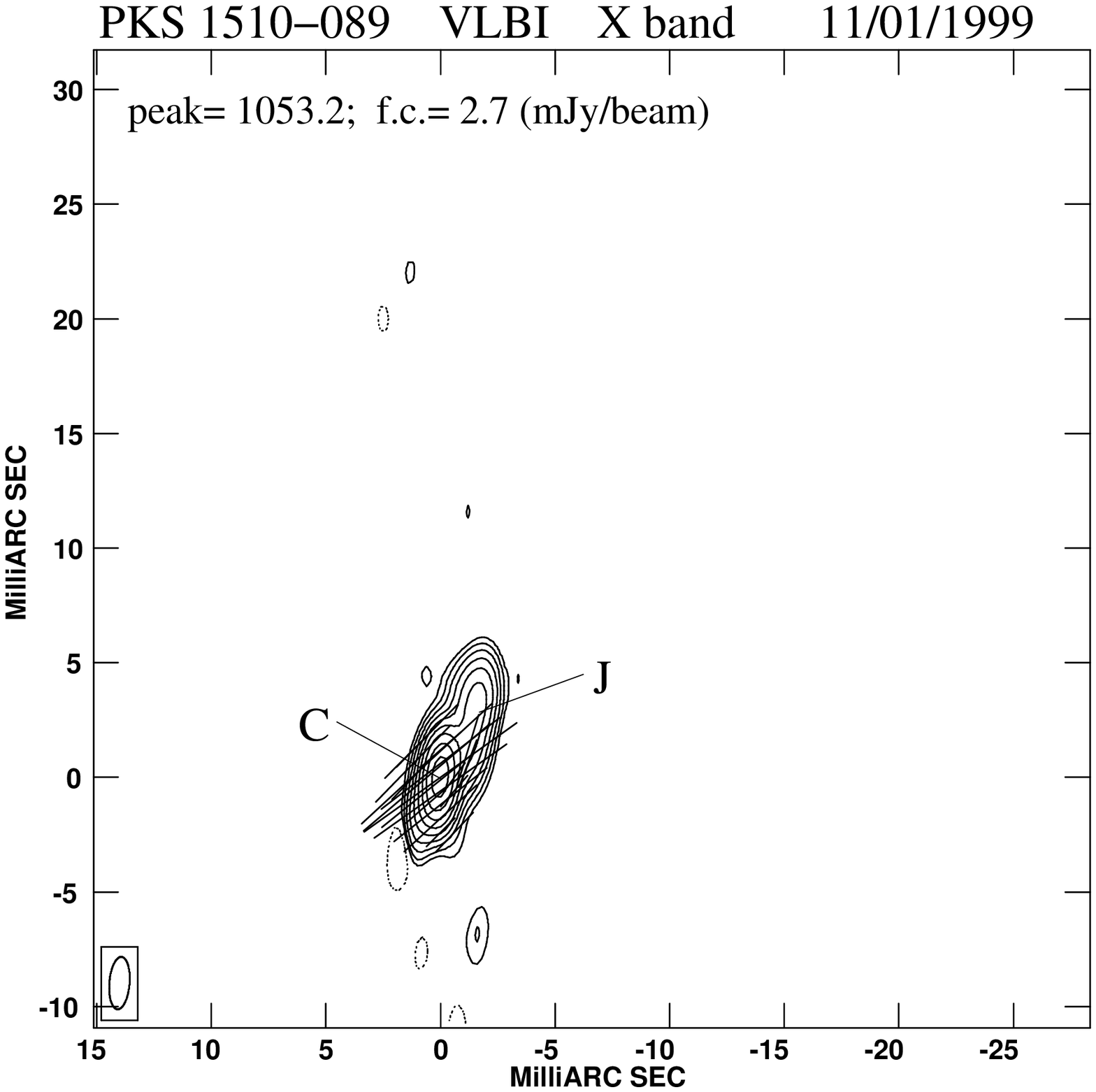}
\includegraphics{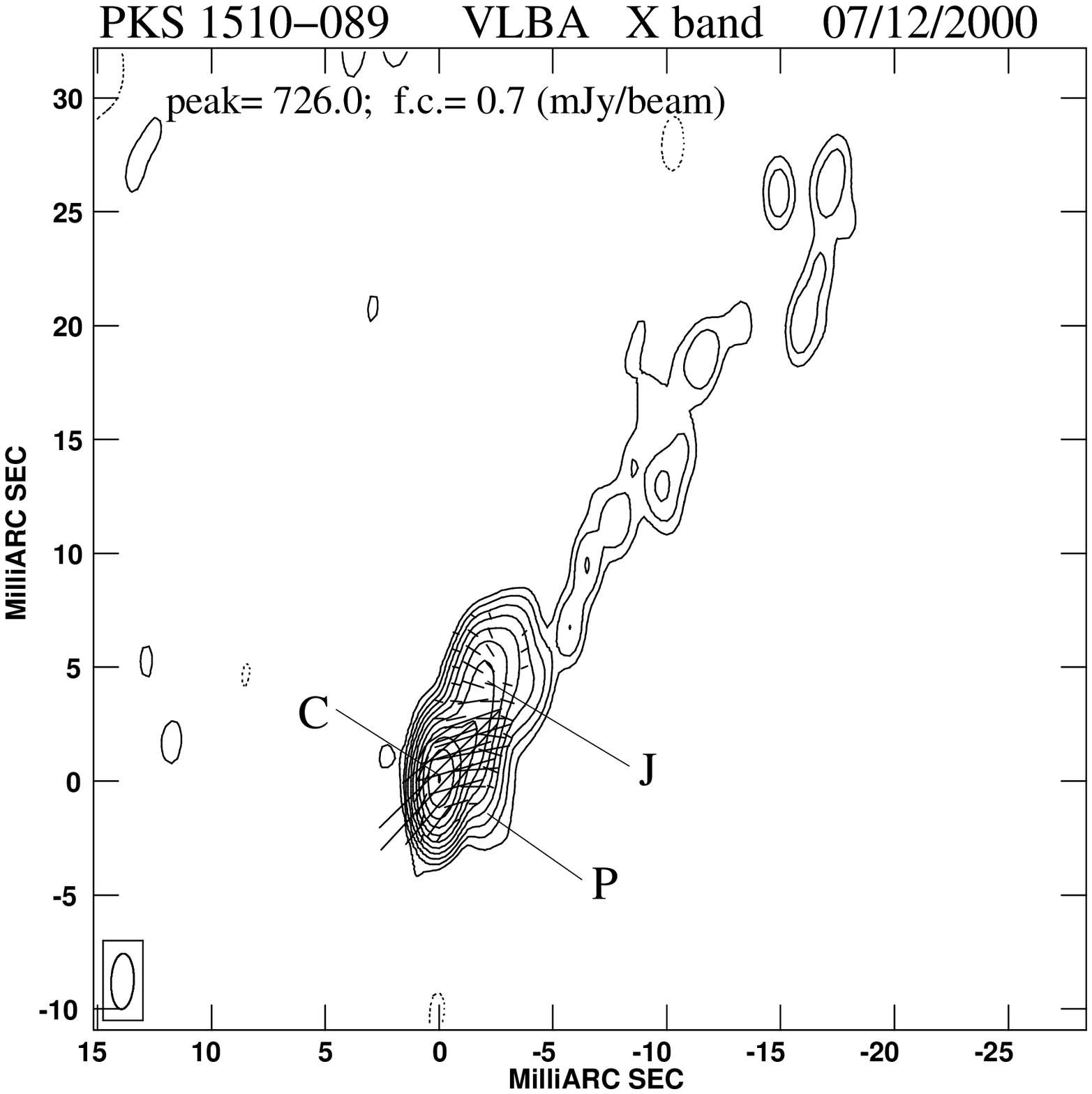}
\includegraphics{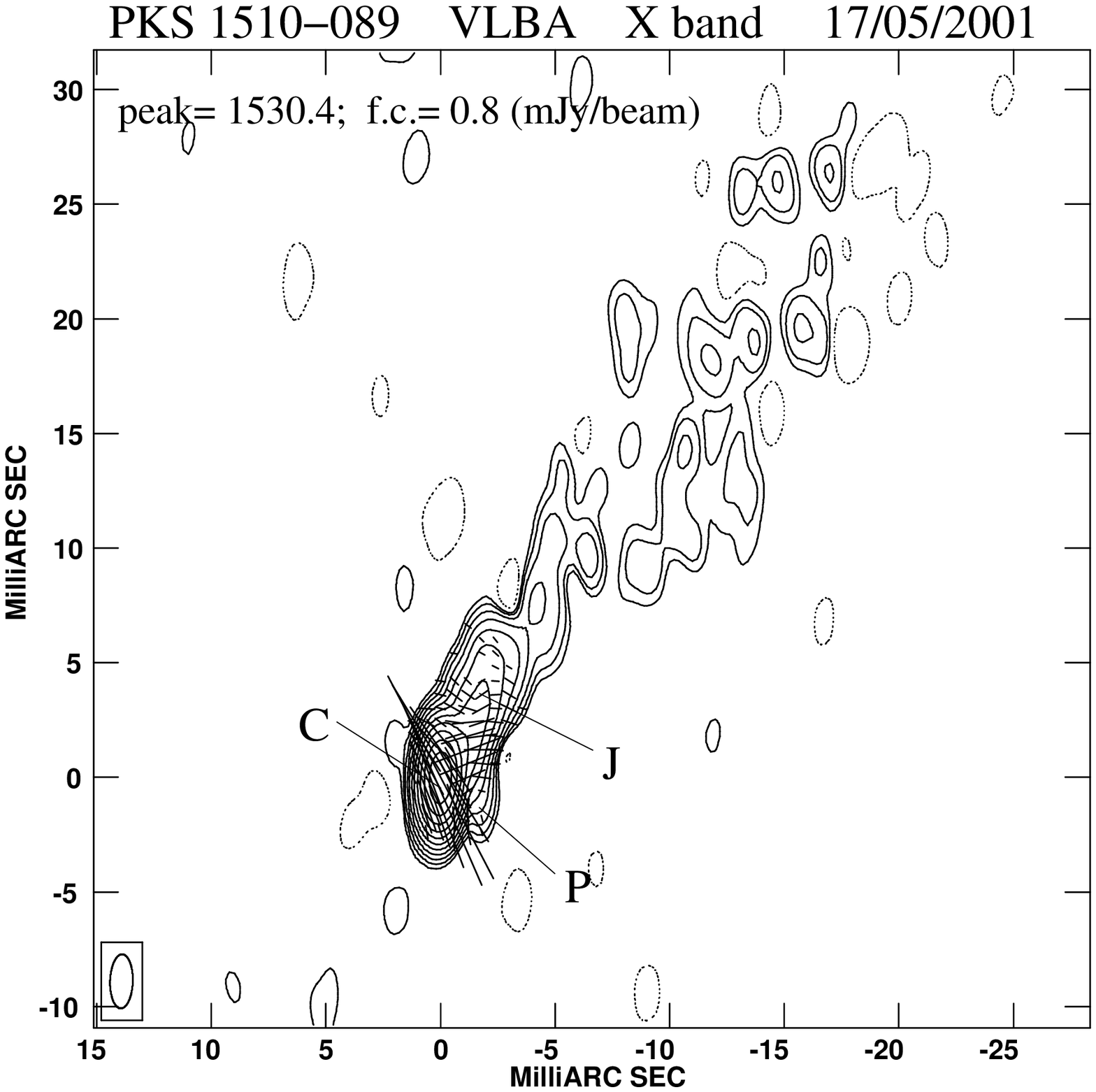}
\includegraphics{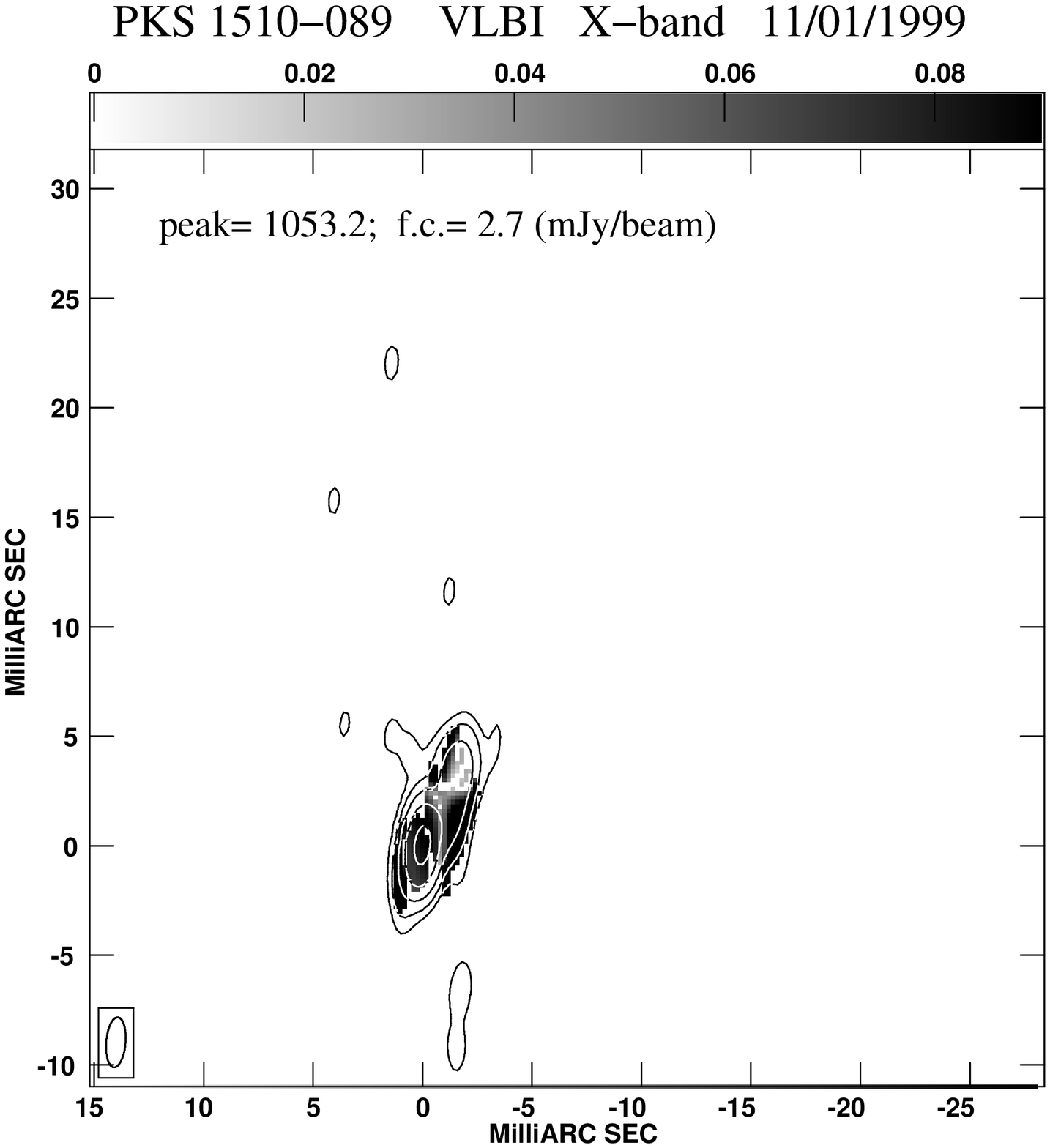}
\includegraphics{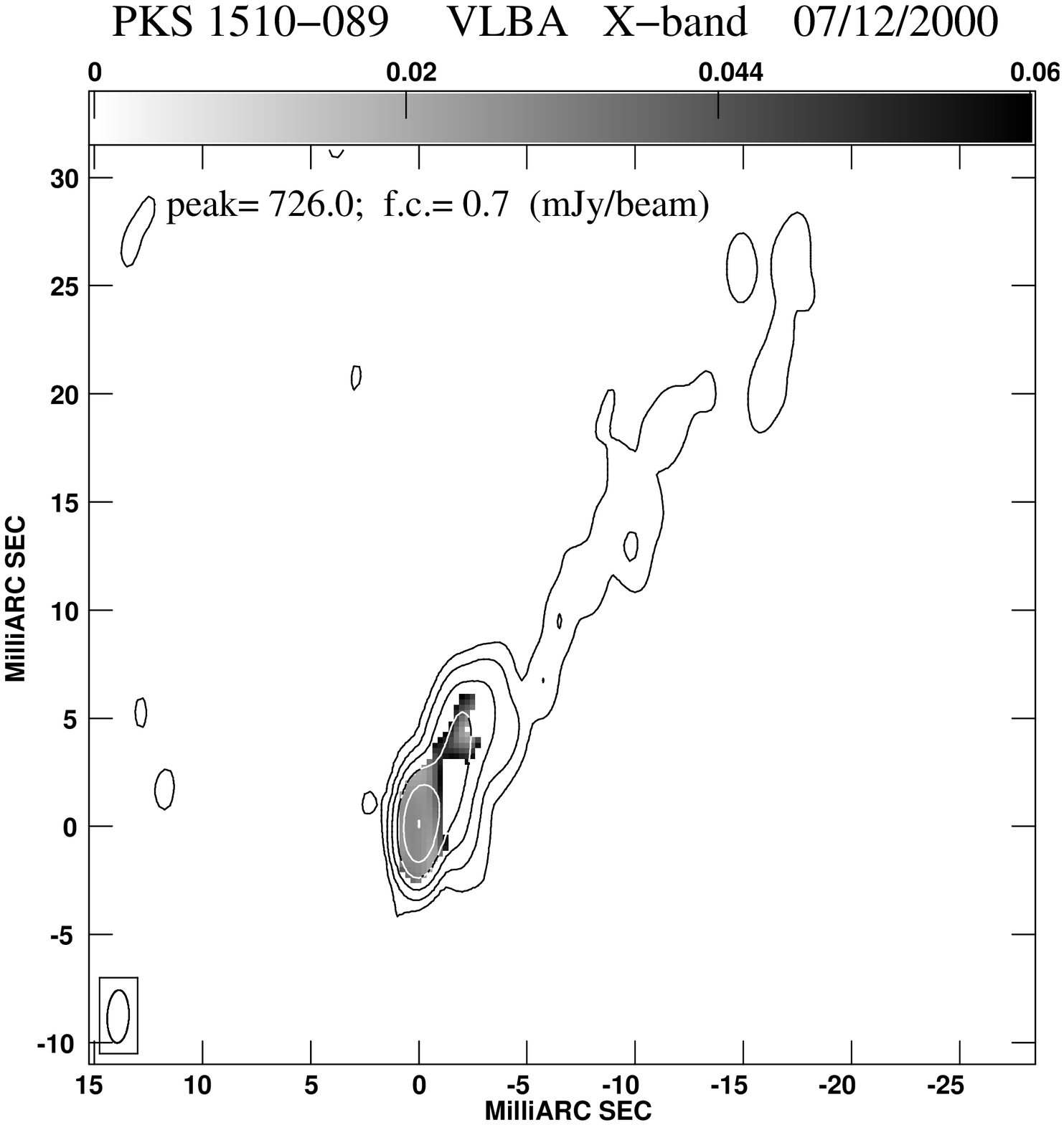}
\includegraphics{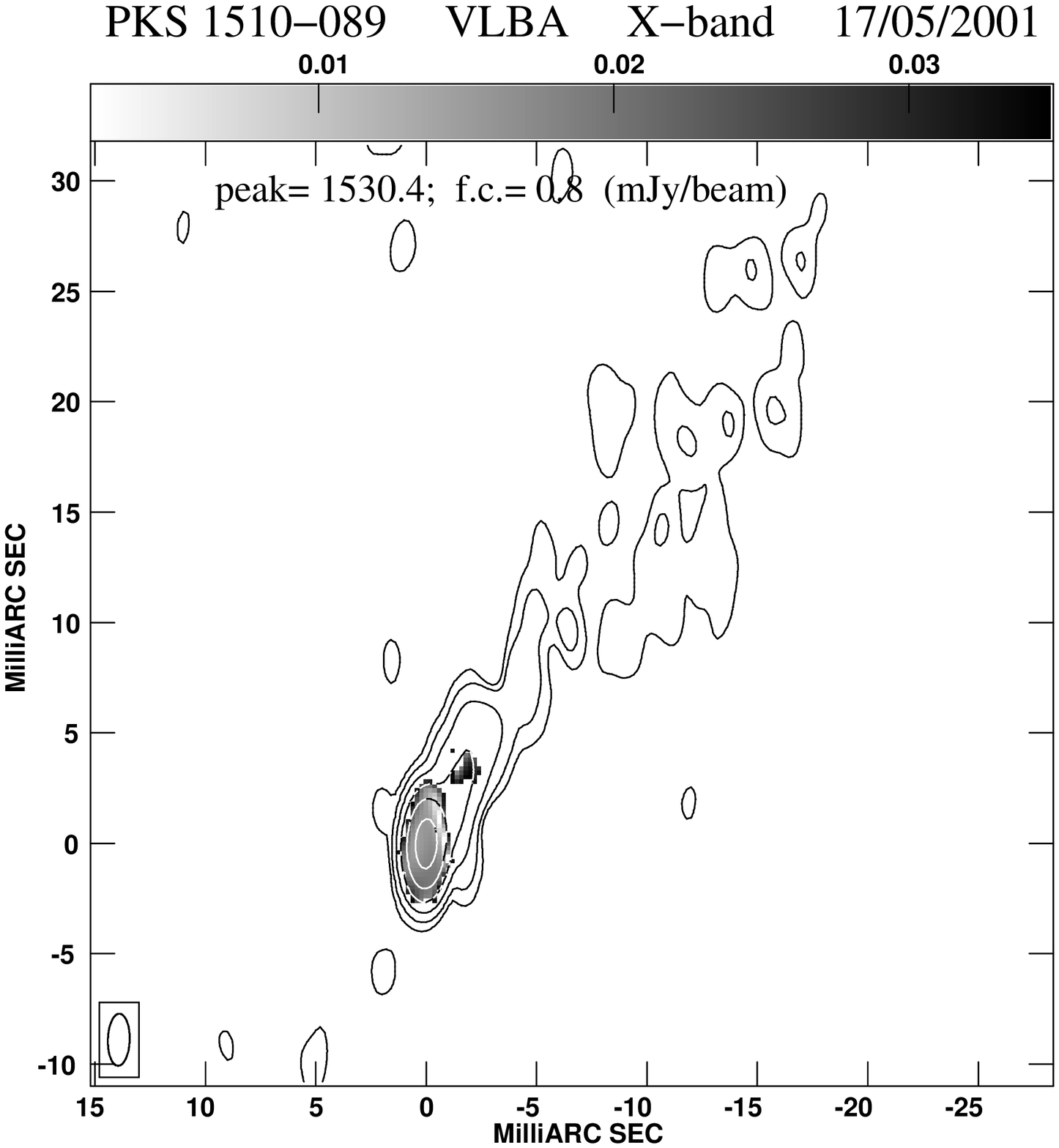}
\vspace{21.7cm}
\caption{VLBI images of the source PKS 1510-089. {\it Upper panel}:
  total intensity VSOP images at 5 GHz and VLBI image at 22 GHz. {\it
    Middle panel}: total intensity VLBA images at 8.4 GHz. {\it
    Bottom panel}: fractional polarization images at 8.4 GHz.
On each
  image we provide the telescope array, the observing band and
  the observing date; 
  the peak flux density in mJy/beam and the first
  contour (f.c.) intensity in mJy/beam, which corresponds to three
  times the off-source noise level. Contour levels increase by a
  factor 2 in the upper and middle panels, and a factor 4 in the bottom
  panel. The restoring beam is plotted on the bottom left
  corner. For the datasets with polarization information, 
the vectors superimposed on the I contours show the position angle of
the $\vec{E}$ vector, where one-millimeter length 
corresponds to 2.5 mJy/beam. The greyscale represents 
the fractional polarization.}  
\label{vlba_images}
\end{center}
\end{figure*}

\begin{figure}
\begin{center}
\includegraphics{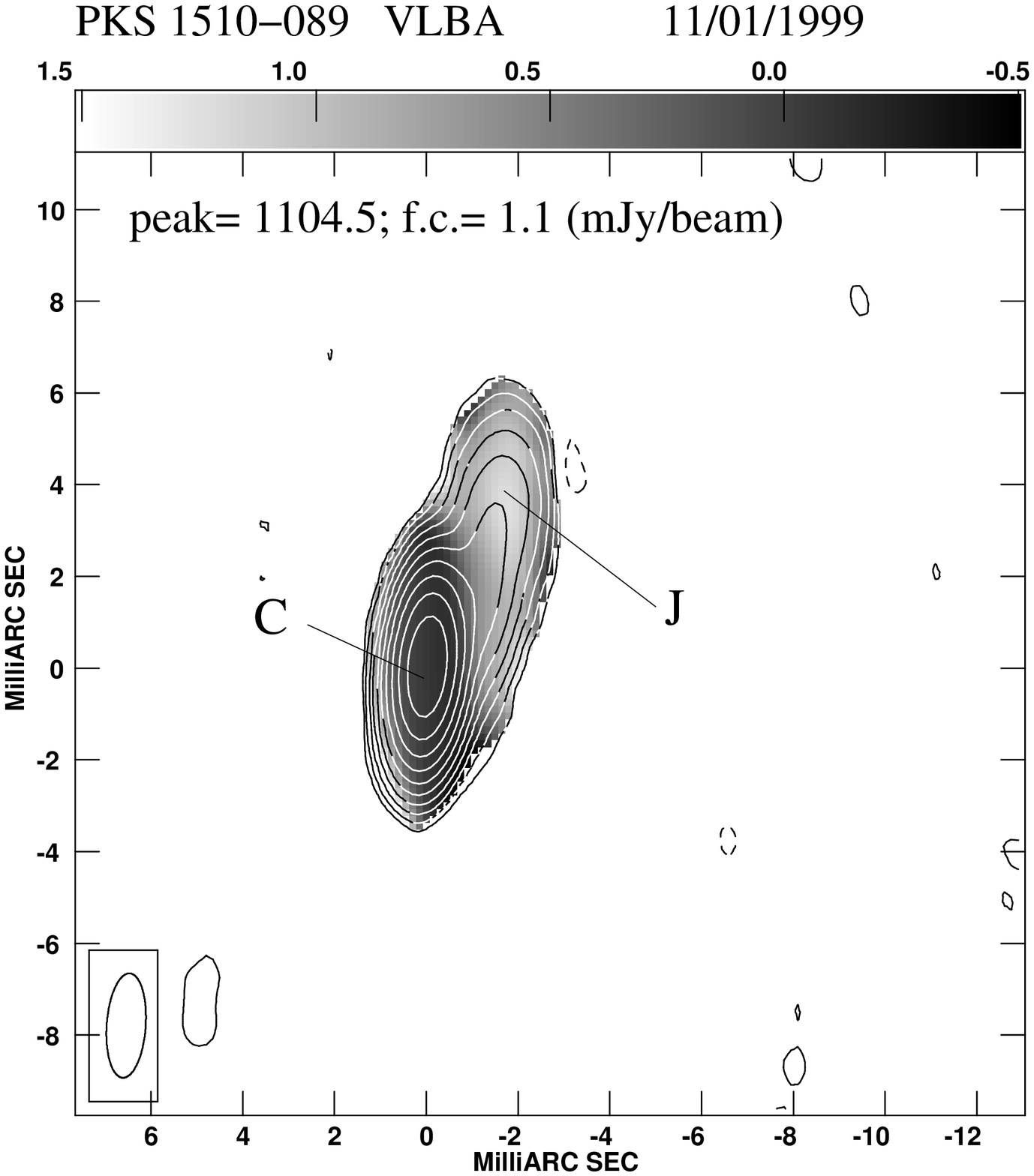}
\vspace{7.5cm}
\caption{Spectral index distribution between 8.4 and 22.2 GHz across
  the target source PKS\,1510-089 superimposed on the 22.2 GHz contours 
  convolved to the 8.4-GHz beam. On the image we provide the peak flux
  density in mJy/beam and the first contour (f.c.) intensity in mJy/beam
  corresponding to 3 times the off-source noise (1$\sigma$). Contour
  levels increase by a factor 2. The restoring beam is plotted on the
  bottom left corner.}
\label{spix}
\end{center}
\end{figure}

Final images for each epoch and at each frequency 
were produced after a number of phase 
self-calibration iterations. Amplitude self-calibration 
was applied at the end of the process using a solution interval
longer than the scan length, to remove residual systematic 
errors and to fine tune the flux density scale, but not to force 
the individual data points to follow the model. The resolution at the
various frequencies are almost comparable (Table \ref{vlba_obs}) 
due to the similar {\it
  uv}-coverage reached with the different interferometers used.
At 8.4 GHz, besides the total intensity (I), images in the Stokes' U and Q 
parameters were produced with the final fully calibrated
datasets.
Final VLBI images at 4.8, 8.4 and 22.2 GHz are shown in
Fig. \ref{vlba_images}.
For the observations carried out on 1999 January 11, we produced also a 
low-resolution image at 22.2 GHz 
using the same {\it uv} range, restoring beam
and image sampling of the 8.4 GHz data in order to produce the
spectral index image, which is presented in Fig. \ref{spix} 
superimposed on the 22.2 GHz contours obtained from the low-resolution
image. 
Information on the VLBI observations is reported in Table \ref{vlba_obs}.\\

\subsection{MOJAVE data}

\begin{figure}
\begin{center}
\includegraphics{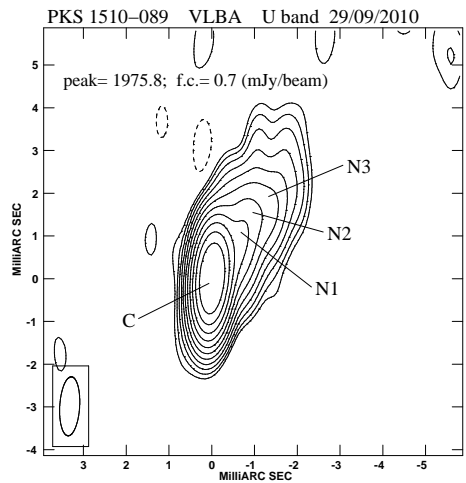}
\vspace{6.5cm}
\caption{An example of the VLBA image at 15 GHz relative to 2010 September
  29. On the 
image we provide the observing band, 
the peak flux density in mJy/beam and the first
  contour (f.c.) intensity in mJy/beam, which corresponds to three
  times the off-source noise level. Contour levels increase by a
  factor 2. The restoring beam is plotted on the bottom left corner. }
\label{mojave_sep10}
\end{center}
\end{figure}

To study the radio variability of PKS\,1510-089, we compared our
observations with multi-epoch 
15-GHz (U band) VLBA data from the MOJAVE programme spanning a time
interval from 1995 July 28 to 2010 December 23. The typical resolution
is about 1.4$\times$0.5 mas.
For each of the 51 epochs analysed, we imported the
calibrated {\it uv}-datasets \citep{lister09a} into the NRAO AIPS package 
and performed a few phase-only self-calibration iterations before
producing the final total intensity images which resulted to be
  fully consistent with those reported by \citet{lister09a}. 
For those datasets in full
polarization mode, we produced also 
Stokes' U and Q images. The rms noise
level measured
on the image plane is in the range of 0.15 and 0.3
mJy/beam. Total intensity images concerning the observing epochs
between July 1995 and July 2008 are published in \citet{lister09a}.
An image of the source structure in September 2010 is presented in
Fig. \ref{mojave_sep10} as an example.\\

\begin{figure}
\begin{center}
\includegraphics{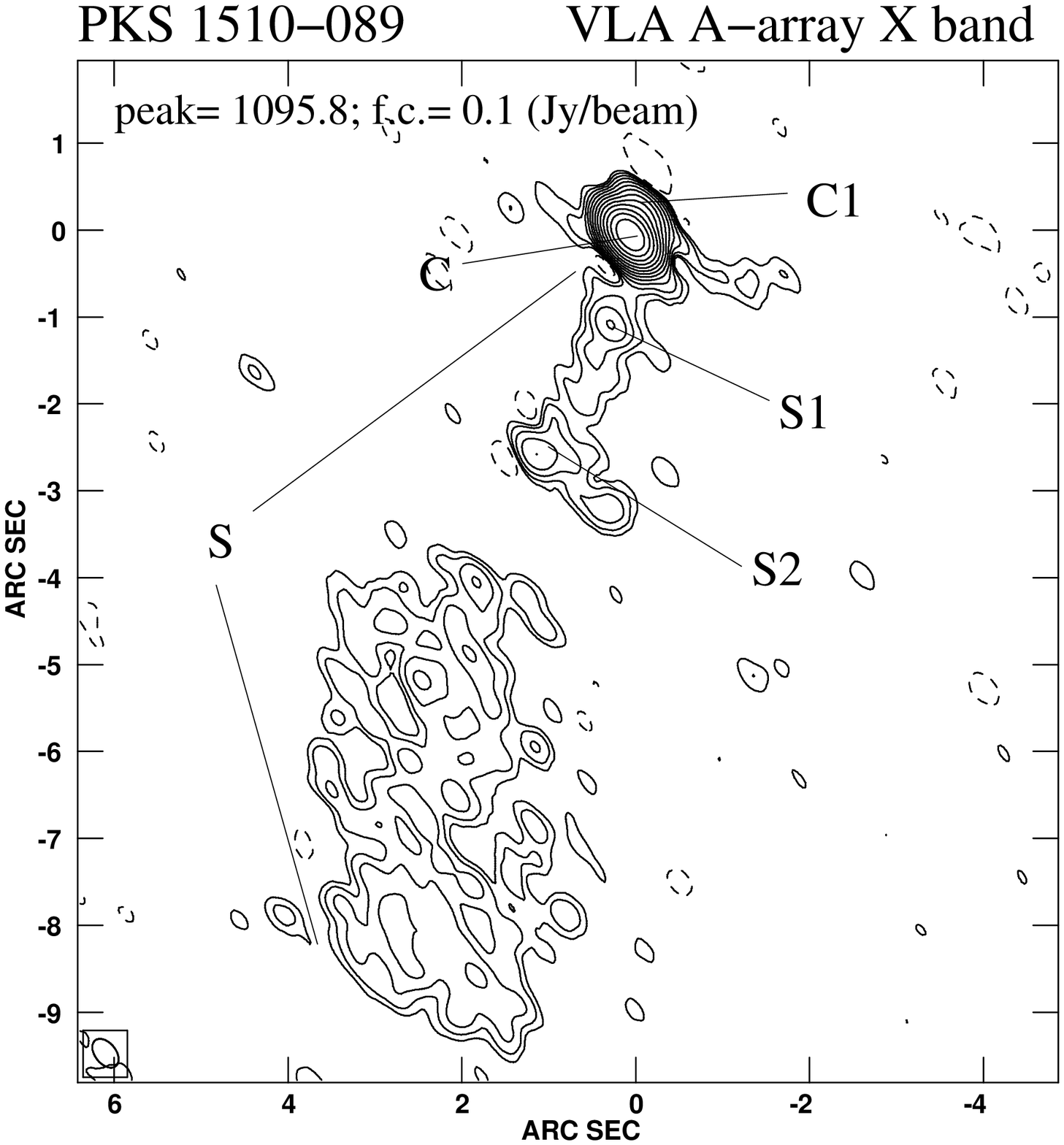}
\vspace{7.0cm}
\caption{VLA A-array image at 8.4 GHz of PKS\,1510-089 
produced from archival data (project AF376) and calibrated
  using the standard procedure developed in the NRAO AIPS package. On the 
image we provide the observing band, 
the peak flux density in mJy/beam and the first
  contour (f.c.) intensity in mJy/beam, which corresponds to three
  times the off-source noise level. Contour levels increase by a
  factor 2. The restoring beam is plotted on the bottom left corner. }
\label{vla_image}
\end{center}
\end{figure}

\subsection{Data analysis}

The flux density and deconvolved angular size of each source component
were measured  
by means of the AIPS task JMFIT, which performs a
Gaussian fit to the source components on the image plane. 
For extended components like the low-surface brightness diffuse
jet visible at 8.4 GHz,
the flux density was derived by
means of TVSTAT, 
which performs an aperture integration over a selected region on the
image plane. When the polarization information was available, we derived
the polarization parameters, such as the linearly polarized flux density,
the fractional polarization, and the (integrated) position angle of
the electric vector $\vec{E}$.
In Table \ref{vlba_table} we report the source parameters concerning 
our proprietary observations at 4.8, 8.4, 22 GHz and 
listed in Table \ref{vlba_obs}. \\
In addition to the data analysis performed on the image plane, a
  further modelfitting with Gaussian components was carried out to the
visibility data at each epoch using the modelfitting option in
Difmap. This approach is preferable in the case we want to derive
small variations in the source structure. Indeed, the analysis of the
visibility data allows us to use the full resolution
capability of the interferometer, without the dependence of the beam
and sampling used in producing the image. Furthermore, it provides
more accurate fit of unresolved structures close to the core
component, as the case of new jet features, thus yielding to a better
determination of the position of each component.\\

\begin{table*}
\caption{VLBI observations of PKS\,1510-089.}
\begin{center}
\begin{tabular}{cllrlll}
\hline
Freq& Code& Obs. date& Dur.& rms& Beam&Array\\
GHz & & &h& mJy/b& mas$\times$mas&  \\
\hline
&&&&&&\\
4.8  & W004&1999 Aug 11& 6 & 0.97&1.80$\times$0.90&VLBA+HALCA\\
4.8  & W004&2000 May 13& 6 & 0.57&1.80$\times$0.90&VLBA+HALCA\\
8.4  &BV027&1999 Jan 11& 1 & 0.90&2.29$\times$0.88&VLBA+EF\\
8.4  &BV042&2000 Dec 7 &1.5& 0.23&2.44$\times$1.00&VLBA\\
8.4  &BV042&2001 May 17&1.5& 0.27&2.36$\times$0.99&VLBA\\
22.2 &BV027&1999 Jan 11&1.5& 0.47&1.70$\times$0.49&VLBA+EF\\ 
&&&&&&\\
\hline
\end{tabular}
\label{vlba_obs}
\end{center}
\end{table*}

\begin{table*}
\caption{VLBI observational parameters of PKS\,1510-089 for the
  observing epochs reported in Table 1.} 
\begin{center}
\begin{tabular}{ccccccccccc}
\hline
Comp&$\nu_{\rm obs}$&$S_{\rm ep1}$&$S_{\rm ep2}$&$S_{\rm
  ep3}$&Pol$_{\rm ep1}$&Pol$_{\rm ep2}$&Pol$_{\rm ep3}$&$\chi_{\rm
  ep1}$&$\chi_{\rm ep2}$&$\chi_{\rm ep3}$\\
    &GHz&mJy&mJy&mJy&mJy (\%)&mJy (\%)&mJy (\%) &deg&deg&deg\\
\hline
&&&&&&&&&&\\
  C& 4.8& 1504& 1161&  -  & -  & - & - & - & - & -\\
  C& 8.4& 1120&  807& 1560& 98 (8.6\%)& 23 (2.8\%)& 26 (1.6\%)&140&110&16\\
  C&22.2& 1139&  -  &  -  & -  &  - &  - & - & - & - \\
\hline
&&&&&&&&&&\\
  J& 4.8&  626&  644&  -  & -  &  - &  - & - & - & - \\
  J& 8.4&  236&  261&  244& 15 (6.3\%)& 13 (5.0\%)& 7 (2.9\%)&70&77&100\\
  J&22.2&  143&  -  &  -  &  - &  - &  - \\
\hline
&&&&&&&&&&\\
Diffuse J&8.4& -   &  34 &   70&  - &  - &  - & - & - & -\\
\hline
\end{tabular}
\label{vlba_table}   
\end{center}
\end{table*}

\section{Results}
\subsection{Morphology}

The radio emission of PKS\,1510-089
is dominated by
the bright core component from which a pc-scale jet emerges at
an angle of 
about -30$^{\circ}$ (Fig. \ref{vlba_images}). 
The pc-scale jet is not straight, but it 
bends at about 2 mas (10
pc) from the core, and
its emission progressively fades until 
5 mas (25 pc) where it goes below the detection limit, in agreement with
images at other frequencies \citep[e.g.][]{homan02b,jorstad05,lister09a}. 
Furthermore, an
additional component almost perpendicular to the jet axis,
visible in several observing epochs and labelled
P in Fig. \ref{vlba_images}, is present
at about 2 mas West from the core. This
component is present only 
in the 15-GHz images with the highest signal-to-noise level 
precluding a reliable 
discussion on its nature. \\
In our deep VLBA observations at 8.4 GHz (Fig. \ref{vlba_images}) 
a low-surface brightness diffuse jet is
visible up to about 25 mas (125 pc) from the core.\\
From the spectral index image (Fig. \ref{spix})
produced comparing the simultaneous observations at
8.4 and 22.2 GHz we find that the core
component has a slightly inverted spectrum ($\alpha$ = -0.02), while
in the jet the
spectral index is 0.5 close to the core and steepens moving outwards.\\
This initial part of the
jet (0.3$^{\prime\prime}$, i.e. 1.5 kpc)
is also visible in the VLA image at 8.4 GHz
(Fig. \ref{vla_image}) as the ``northern component'' C1, in the
opposite direction of the kpc-scale structure (labelled S). The large
scale structure is roughly collimated up to
a distance of about 2 arcsecond (10 kpc) from the core. Then it
slightly bends to the West,  
forming an extended low-surface brightness lobe-like
feature. 
Two compact regions (S1 and S2 in
Fig. \ref{vla_image}), likely jet knots,  are located at
1.0$^{\prime\prime}$ ($\sim$ 5 kpc) and 2.7$^{\prime\prime}$ ($\sim$13
kpc) from the core. \\

\subsection{Flux density and polarization}

\begin{figure}
\begin{center}
\includegraphics{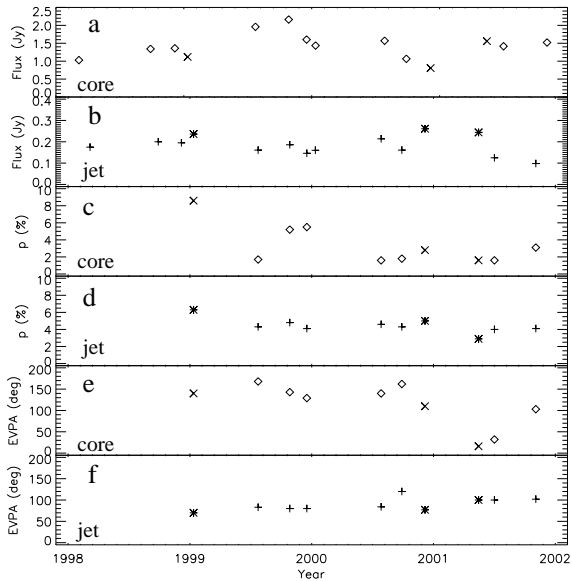}
\vspace{8cm}
\caption{Total intensity and polarization properties at 8.4 and 15 GHz
  of PKS\,1510-089. Core parameters are marked by a diamond at
  15 GHz, and by a cross at 8.4 GHz. Jet parameters are marked by a
  plus sign at 15 GHz and by an asterisk at 8.4 GHz.
  {\it Panel {\bf a}}: total intensity flux density 
of the core; {\it panel {\bf b}} total intensity flux density of the 
jet; {\it panel {\bf c}}: fractional
  polarization of the core component; {\it panel {\bf d}}: fractional
  polarization of the jet component; {\it panel {\bf e}}: position
  angle of the electric vector (EVPA) of the core component; 
{\it panel {\bf f}}: EVPA of the jet component. 
  Uncertainties are within the symbols.}
\label{plot_mojave99}
\end{center}
\end{figure}

The analysis of the radio lightcurves of PKS\,1510-089 shows
strong flux density variability, where low states alternate with
flares \citep{venturi01,jorstad01,hughes92}. 
Sometimes additional episodes of abrupt flux
density increase have been registered. These studies are based on
single-dish observations where it is not possible to
disentangle the contribution of the jet from the one arising from the
core, and changes in the parsec-scale structure may be washed out by
the contribution of the stationary components. 
The availability of high-resolution VLBI data allows us to
separate the core and jet emission and thus to determine how the flux
density changes in both components. Comparing our three-epoch data at
8.4 GHz (Table \ref{vlba_obs}) 
we find that the core flux density varies of about 50\%
between 1999 and 2001, while the scatter in the jet flux density 
is within 10\%. We compare the flux density lightcurve at 8.4 GHz 
with the 15-GHz VLBA data
from the MOJAVE programme \citep{lister09a} obtained between 1998 and
2001. 
In Figs. \ref{plot_mojave99}a,b we report
the flux density of the core and jet components. The core flux density
at 15 GHz
clearly shows strong variability reaching a maximum in October 1999,
when it has almost doubled its value, then followed by a clear decrease
lasting till the end of 2001 (Fig. \ref{plot_mojave99}a). 
After this period the core flux
density increases again, in agreement with the trend found
at 8.4 GHz, but the observations are quite sparse precluding a
more detailed study of the core variability. On the other hand, the
jet flux density at 15 GHz varies about 25\% till the end of 2000, 
while the scatter increases in 2001 (Fig. \ref{plot_mojave99}b). 
The jet flux density has more
moderate variability than the core. Despite the lack of
simultaneous multifrequency observations, we can confirm
a steep overall spectral index for the jet ($\alpha =$0.5 - 0.7)
between 8.4 and 15 GHz. On
the other hand a characteristic value of the spectral index of the
core could not be determined due to its strong and irregular
variability associated with changes in the opacity.
A realistic value for the core spectral index could be
derived only for the observations carried out on January 1999 
when simultaneous 
observations at 8.4 and 22 GHz are available, providing a flat
spectral index (Fig. \ref{spix}).\\
The polarization properties of both core (Fig. \ref{plot_mojave99}c,e) 
and jet (Fig. \ref{plot_mojave99}d,f) components are quite
different. 
In the three observing epochs at 8.4 GHz we find that both the
core and the jet are polarized with a fractional polarization varying
between 1.6 and 8.6 per cent in the core region, and about 2.9 and 6.3 per
cent in the jet (Table \ref{vlba_table}, bottom panel of
  Fig. \ref{vlba_images}, and Fig. \ref{plot_mojave99}c,d). 
At 15 GHz the polarization percentage of the core is
very variable, between 1.6 and 5.5 per cent
(Fig. \ref{plot_mojave99}c), and shows no evident correlation with
the total intensity flux density. On the other hand, the
polarization percentage of the jet at 15 GHz (Fig. \ref{plot_mojave99}d)
is roughly constant around 4.5 per cent.
In the core component the position angle of the
electric vector $\vec{E}$ (EVPA) at 8.4 GHz is 
140$^{\circ}$ and 110$^{\circ}$ during the first two epochs, then it
changes abruptly in May 2001, when it is
16$^{\circ}$. A similar behaviour is found at 15 GHz, where the EVPA is
between 168 and 130 degrees from 1999 to 2000, i.e. roughly parallel
to the jet axis, while in June 2001
it is 32 degrees (Fig. \ref{plot_mojave99}e),
becoming perpendicular to the jet direction. On the other hand,
  the jet EVPA does not show such a large variation, being between 80
  and 100 degrees, with the exception of September 2000 when it
  turned out to be 120 degrees. We note that 
only the integrated rotation
measure (RM) is available for this source (-15$\pm$1 rad m$^{-2}$,
Simard-Normandin et al. 1984). The corrections to the observed EVPA to
obtain the intrinsic orientation are therefore negligible.\\

\subsection{Proper motion}

\begin{figure}
\begin{center}
\includegraphics{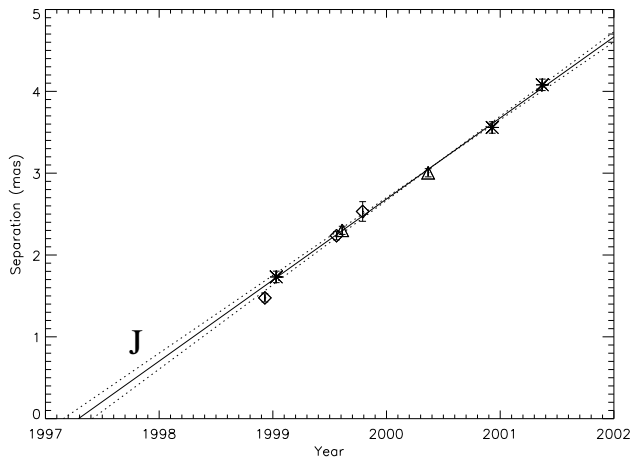}
\vspace{6cm}
\caption{Changes in separation with time between components C and
  J. Asterisks, diamonds, and triangles  refer to 8.4-GHz VLBA data,
  15-GHz VLBA data, and 4.8-GHz Space-VLBI (VSOP) data respectively.
The solid line represents the regression fit to the 8.4-GHz VLBI
  data, while the dashed-lines represent the 
uncertainties from the fit parameters. Error bars are determined as in
Section 3.3.}
\label{fit_vsop}
\end{center}
\end{figure}

\begin{figure}
\begin{center}
\includegraphics{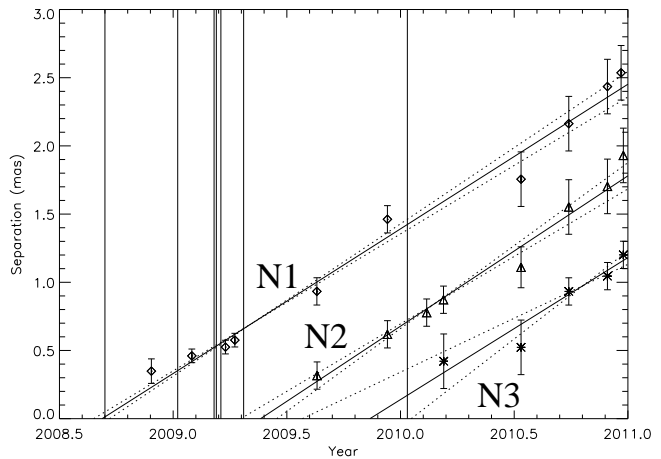}
\vspace{6cm}
\caption{Changes in separation with time between components C and
  N1, N2, and N3. 
The solid line represents the regression fit to the data, while the
dashed-lines represent the uncertainties from the fit parameters. Error
  bars are determined as in Section 3.3. 
Vertical lines show
  $\gamma$-ray flares \citep{pucella08, dammando09a,
    ciprini09, dammando09b, pucella09, vercellone09,
    cutini09,striani10}. }
\label{fit_mojave}
\end{center}
\end{figure}

The multi-epoch analysis of the pc-scale morphology of PKS\,1510-089 
shows a considerable 
evolution of the source structure: jet components emerge from the core 
at different
times and their changes can be followed by comparing observations carried
out after short time intervals. 
With the aim of characterizing 
variations in the source structure we modelfitted 
the visibility data at each epoch (see Section 2.3). 
Direct comparison of models obtained independently at each
epoch is not the best approach to detect small changes 
\citep{conway92}. For this reason, we produced a
zero-order model consisting of 3 elliptical Gaussian components,
which was used as the initial model in modelfitting 
the visibility data of each observing epoch. 
Errors $\Delta r$ associated with the component position were
estimated by means of:\\ 

\begin{equation}
\Delta r = a /(S_{\rm p}/{\rm rms})\\
\label{eq_rms}
\end{equation}

\noindent where $a$ is the component deconvolved major axis, 
$S_{\rm p}$ is its peak flux
density and rms is the 1$\sigma$ noise level measured
on the image plane \citep{polatidis03}. In the case the errors estimated
by Eq. \ref{eq_rms} are unreliably small, we assume a more
conservative value for $\Delta r$ that is 10\% of the beam.
The data points are then fitted by a linear model that minimizes the
chi-square error statistic. \\
The linear fit on the three epochs of 8.4-GHz data (Table \ref{vlba_obs} and
Fig. \ref{vlba_images}) provides an angular
separation rate of 0.990$\pm$0.040 mas/yr that corresponds to
$\beta_{\rm J}=16.2\pm0.7$. 
From the linear
back extrapolation fit, and considering the uncertainties on the fit parameters,
we estimate the time of zero separation $T_{0}$
between the jet component and
the core, that results to be $T_{\rm 0,J}=$1997.42$\pm$0.12.
The accuracy
of the fit has been tested also by comparing
the component separation derived by modelfitting the visibility of
the two epochs 4.8-GHz Space-VLBI data and the three epochs at 15 GHz
available from the MOJAVE project in the same
time interval (left panel in Fig. \ref{fit_vsop}), all with a
similar resolution. \\
To extend the analysis of the knot separation speed till the end of
2010, 
we modelfitted the data at 15 GHz from the MOJAVE
programme. Results from the analysis of data between 1995 and 2008 were
already published in previous works by \citet{homan01} and \citet{lister09b},
where the separation velocities found for the detected knots range between
15$c$ and 20$c$\footnote{In the Appendix, we 
re-modelfitted these datasets in order to
  compare results obtained with a consistent approach.}.
Since 2007, when also {\it Fermi} and AGILE satellites could provide
crucial information to relate the radio and high-energy $\gamma$-ray 
emission, we
could follow the evolution of three additional knots, labelled N1, N2 and N3 in
Fig. \ref{mojave_sep10}. For N1, a 1$\sigma$ separation speed of 675
$\mu$as/yr (i.e. 11.6$c$) was reported by \citet{abdo10b} using data
between September 2008 and December 2009. In our analysis
we expand the time baseline to December 2010.
We determine
the angular separation speeds of these three new knots from the core, 
which we assume stationary: we derive 
1.060$\pm$0.056 mas/yr, 1.102$\pm$0.113 mas/yr, and
1.041$\pm$0.250 mas/yr for N1, N2 and N3 respectively, 
which correspond to
$\beta_{\rm N1}=17.3\pm 0.9$, $\beta_{\rm N2}=18.0\pm 1.9$,
$\beta_{\rm N3}=17.0\pm 4.0$. For each knot the time of
zero separation obtained from the fit is $T_{\rm
  0,N1}=2008.61\pm 0.11$, $T_{\rm 0,N2}=2009.38\pm 0.17$, and $T_{\rm
  0,N3}= 2009.93\pm 0.25$ (Fig. \ref{fit_mojave}). 
Components 
N1 and N2 were also detected in VLBA observations at 43 GHz by
\citet{marscher10}. Given the lower spatial resolution (and larger
opacity) with respect to the 43 GHz, these components were
detected at 15 GHz a few months later when their separation from the
core was large enough to be resolved. The apparent separation speeds
derived at 43 GHz are 24$\pm$2 and 21.6$\pm$0.6 for N1
and N2 respectively, i.e. systematically larger than our values. This
discrepancy may be due to the lower resolution of the 15-GHz data that
may cause a blending with other features, as also suggested in
\citet{abdo10b}. However, it must be noted that the higher 
  separation speed found at 43 GHz is not reflected in a substantially
  different origin time of the jet components. In fact, in the case
  of N1 and N2 components the time of zero separation derived from
  43-GHz data are 2008.57$\pm$0.05 and 2009.34$\pm$0.01 respectively
  \citep{marscher10}, i.e. within the errors estimated from the 15-GHz data.
In Table \ref{tabella_moto} we report the apparent
separation speed and the ejection date of the superluminal knots
of PKS\,1510-089.\\

\begin{table}
\caption{Proper motion of the jet components of PKS\,1510-089. 
Column 1: source component;
Col. 2: number of fitted epochs; Col. 3: apparent
speed; Col. 4: zero-separation time; Col. 5: zero-separation time in
Julian Date (JD); Col. 6: Reference: 1: this paper; 2:
\citet{homan01}; 3: \citet{lister09b}; 4: \citet{abdo10b}; 5:
\citet{marscher10}.}
\begin{center}
\begin{tabular}{lrcccc}
\hline
Comp&N$_{\rm ep}$&$\beta_{\rm app}$&$T_{0}$&$T_{0}$ (JD)&Ref.\\
\hline
&&&&&\\
J&8&16.2$\pm$0.7&1997.42$\pm$0.12&2450602$\pm$45 &1\\
A&11&15.0$\pm$0.7& 1994.92$\pm$0.18&2449689$\pm$66&1,2,3\\
B&19&18.6$\pm$0.5&2005.42$\pm$0.11&2453524$\pm$40 &1,3\\
N1&10&17.3$\pm$0.9&2008.61$\pm$0.11&2454689$\pm$36 &1,4,5\\
N2&8&18.0$\pm$1.9&2009.38$\pm$0.17&2454971$\pm$62 &1,5\\
N3&5&17.0$\pm$4.0&2009.93$\pm$0.25&2455171$\pm$91 &1\\
C&10&20.2$\pm$1.2&1997.85$\pm$0.20&2450759$\pm$73&3\\
D&8&18.9$\pm$1.3&1999.10$\pm$0.25&2451215$\pm$91&3\\
E&7&14.8$\pm$0.9&2000.20$\pm$0.20&2451617$\pm$73&3\\
F&6&19.1$\pm$1.5&2004.42$\pm$0.28&2453129$\pm$73&3\\
&&&&&\\
\hline
\end{tabular}
\label{tabella_moto}
\end{center}
\end{table}

\section{Discussion}

The population of blazar objects represents a sub-class of AGN
characterized by high levels of flux density variability across the entire
electromagnetic spectrum. In the radio band the increase in
luminosity is first detected at high frequencies, at millimeter
wavelengths, subsequently followed at lower frequencies with some delay, 
consistent with opacity effects. 
Episodes of enhanced
radio luminosity also seem to be related to changes in the pc-scale radio
structure where new components are ejected with apparent superluminal
velocity \citep[e.g.][]{wagner95,jorstad01}.\\
In the following we discuss our results on the multi-epoch analysis of 
the total intensity and polarimetric radio properties, and we compare the
source activity at various frequency ranges.\\ 

\subsection{Superluminal motion and polarization analysis}

\begin{figure}
\begin{center}
\includegraphics{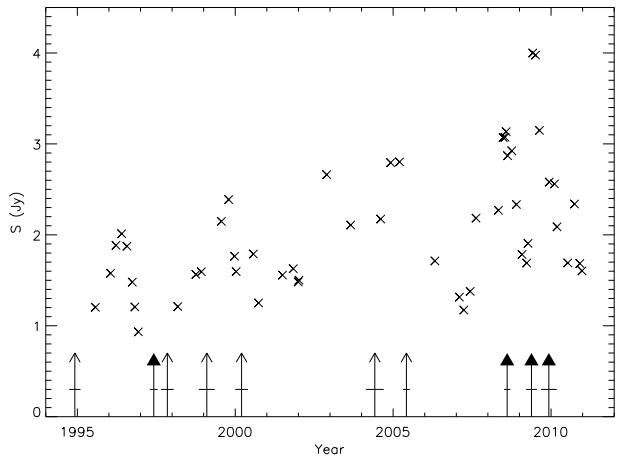}
\vspace{6cm}
\caption{The 15-GHz lightcurve from VLBA MOJAVE data. The
  uncertainties on the flux density are within the symbols.
Thin arrows represent the fitted ejected date of
  superluminal components from \citet{lister09b}, while thick arrows
  refer to the jet components studied in this paper. Each arrow
    refers to a different superluminal component. The horizontal bars
    indicate the uncertainty on the ejection date.}
\label{1510_arrow}
\end{center}
\end{figure}

By means of the multi-epoch analysis of the parsec-scale structure of
PKS\,1510-089, various knots emerging from the
core at different times have been detected. In particular, we obtain
a solid estimate for one of the components detected between
1999 and 2001, thanks to the addition of 5 new position measurements
at 4.8 and 8.4 GHz.  
Interestingly, comparing
the properties of the knots analysed in this paper
with those studied in previous works (Table \ref{tabella_moto})
we found that all the 
knots are separating from the core with an apparent velocity
that ranges between 15$c$ and 20$c$, and all the jet components are
well aligned along the same position angle, suggesting that
no jet precession is
taking place on a time lag longer than a decade in our frame. \\
In Fig. \ref{1510_arrow}, we report the time of zero separation of all
the components ejected between 1994 and 2010, either analysed in this
paper, or from the literature. The ejection of a new component
likely occurs roughly once per year. 
The gaps between 2000 and 2004 is likely due to
sparse observations available, precluding to reliably follow the
evolution of the source components in this period.  \\
Since 1995, in addition to the typical variability interchanging
low-activity with high-activity states, the source shows on average a
slight increase of the total flux density
(Fig. \ref{1510_arrow}). However, the sparse distribution of the
data points does not allow us to unambiguously relate the ejection
date of the superluminal components, represented by the arrows in Fig.
\ref{1510_arrow}, with the various states of the source variability.\\
From the analysis of the parsec-scale resolution data we could
construct the lightcurves for the core and jet components separately,
and to study a connection between the total intensity and the
polarized emission. 
The core has high
degree of variability, both in the total intensity flux density and in
polarization properties, but the data points are too sparse to compare 
changes in the total intensity and polarized
emission throughout the entire time interval. 
An interesting case is represented by the
simultaneous enhancement of the total intensity flux density and the
decrease of the fractional polarization observed in the second half of 1999, 
and the rotation
of EVPA around May-July 2001, with a swing of 85$^{\circ}$ and
130$^{\circ}$ at 8.4 and 15 GHz respectively
(Fig. \ref{plot_mojave99}). 
Possible explanations may be related either to variation in the
  opacity in a newly ejected self-absorbed component,
or to a highly ordered magnetic field produced by the
compression of tangled magnetic fields by shocks. In
the former scenario the evolution of the jet knot implies a change
between the opacity regimes. As a
consequence both the total intensity and the polarized 
flux densities should decrease during the transition, 
whereas the magnetic field should jump by 90 degrees. As the
opacity decreases, the total flux density should increase in the
optically-thick part of the spectrum, while the polarization
percentage decreases. As the emitting region expands becoming optically-thin,
the radio emission decreases, while the 
fractional polarization increases and
the magnetic field flips of 90 degrees, going back to its original
value. It must be noted that 
changes of the EVPA of 90$^{\circ}$ are expected from opacity effects
during the transition between the two regimes. Since the
process is quite fast, a dense time sampling of the polarization
information (as the one obtained between 2007 and 2010, Fig. \ref{multi_chi})
is crucial to detect the transition between the regimes. A sparser
sampling, like the one between 1999 and 2001
(Fig. \ref{plot_mojave99}), may provide EVPA changes larger/smaller
than the expected 90$^{\circ}$.
The transition between the optically-thick and -thin
  regimes occurs in the presence of high opacity values (e.g. $\tau
  \sim$ 5 - 10, see for example Aller (1970) and Pacholczyk (1970)
  for a detailed discussion). Such
  opacity would cause a dramatic drop of the total intensity flux
  density during the transition. The lack of observations during the
  first half of 1999, i.e. when the new jet component should have
  originated, do not allow us to unambiguously confirm the opacity
  scenario. However, the increase of the total flux
    density observed between 1999 and 2000 is difficult to explain in
    the presence of such high values of the opacity, making this
    scenario unlikely.\\ 
The alternative scenario assumes the formation of a shock that causes
the compression of the plasma along the 
propagation axis. This generates the amplification of the perpendicular
component of the magnetic field with respect to the parallel one, and
an enhancement of the luminosity. This scenario would predict
an increase of the fractional polarization instead of 
the drop observed close to the radio outburst.
 However, we must note that if the shock is oblique instead of
 transverse, the expected variations in the polarization properties
 are different, 
 and strongly related to the obliqueness of the shock itself, 
and to the characteristics of the underling magnetic field like its
order and strength. Therefore, a reliable interpretation of the
observed polarization trends requires a much better sampling than that
available between 1998-2001 (Fig. \ref{plot_mojave99}), leaving the debate on
the nature of the main mechanism at work still open.\\
The analysis of the
lightcurves and polarization trends between 1998 and 2001, shown in
Fig. \ref{plot_mojave99}, suggests that the variations of the 
total intensity flux density and the polarization properties may be
explained by both the previous scenarios. Indeed, the total intensity
and polarization properties can be 
related to the evolution (likely adiabatic expansion) of either a new
jet component or a shock originated at the beginning of 1999. Support to this
interpretation comes from the detection of a superluminal component  
that possibly originated in 1999.10$\pm$0.25 \citep{lister09b}. The
rotation of the angle found in 2001 may be explained considering that
this new component moves far enough from the core to be
resolved and the intrinsic polarized emission of the core can be
separated from that of the jet.\\

\subsection{Multifrequency analysis}

\begin{figure}
\begin{center}
\includegraphics{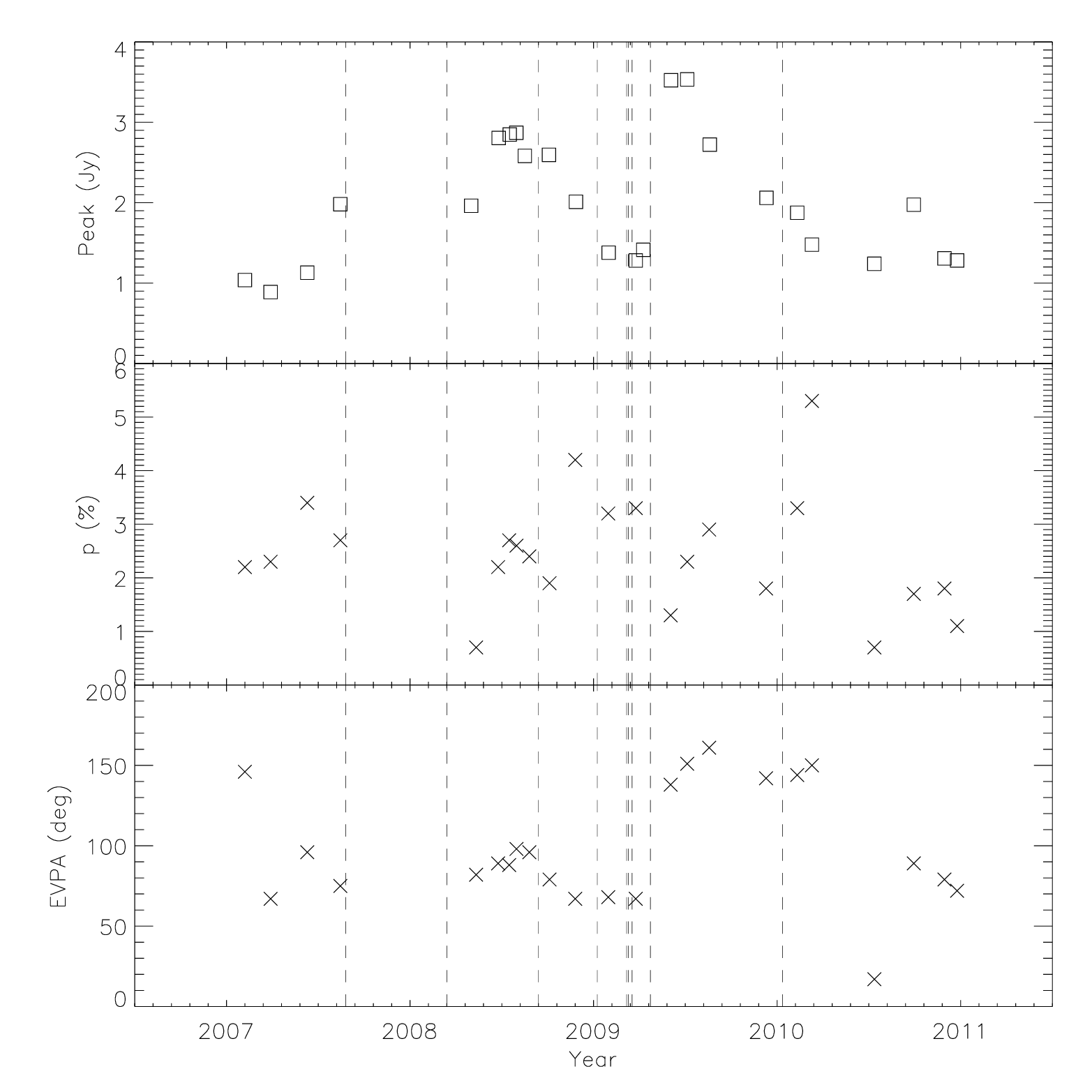}
\vspace{7.5cm}
\caption{Multi-epoch 15-GHz VLBA peak flux density, restored with
  the same beam size ({\it upper panel}), fractional polarization 
({\it middle panel}), and
  polarization angle ({\it lower panel}) of the core component of
  PKS\,1510-089. Uncertainties are within the symbols. 
  Vertical lines are relative to high-state
  $\gamma$-ray activity \citep{pucella08, dammando09a,
    ciprini09, dammando09b, pucella09, vercellone09,
    cutini09,striani10}.}
\label{multi_chi}
\end{center}
\end{figure}

\begin{figure}
\begin{center}
\includegraphics{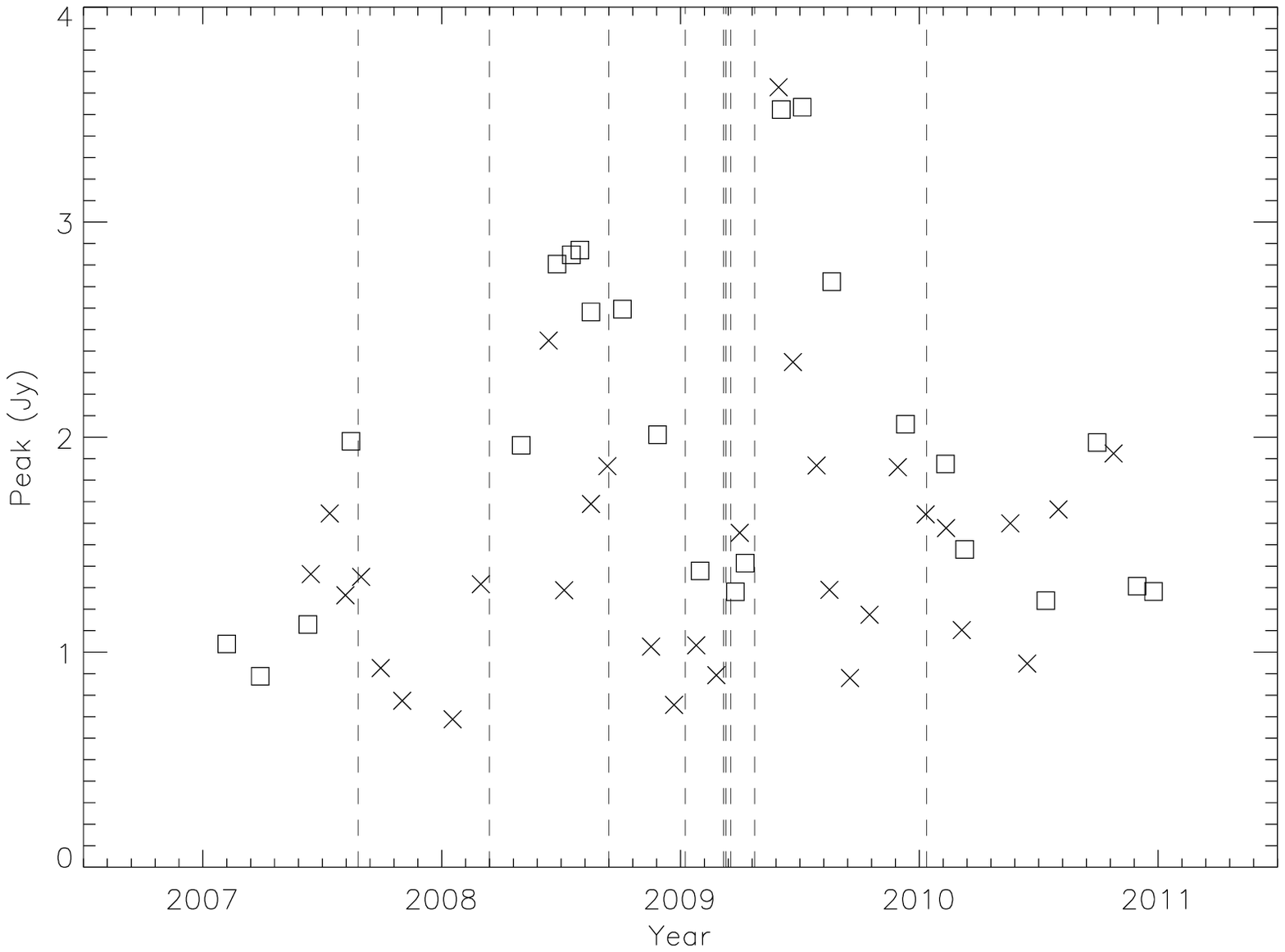}
\vspace{6.5cm}
\caption{Multi-epoch peak flux density at 15 GHz ({\it squares}) and
  43 GHz ({\it crosses}) of the central component of
  PKS\,1510-089. Vertical lines are relative to high-state
  $\gamma$-ray activity \citep{pucella08, dammando09a,
    ciprini09, dammando09b, pucella09, vercellone09,
    cutini09,striani10}. }
\label{peak_flux}
\end{center}
\end{figure}

The lightcurves considered in the above 
explanation would require a more frequent sampling and for this reason 
a definitive interpretation of the physical mechanisms
at work is precluded. A similar consideration may apply to explain the source
behaviour around April 2009, when information
at other wavelengths \citep[e.g.][]{abdo10b,marscher10} is available.
The multiwavelength lightcurve presented by \citet{abdo10b} shows a
flux density enhancement around that period that is first detected at
230 GHz, and after some delay at lower frequencies. The
same behaviour is visible in Fig. \ref{peak_flux} where the 
lightcurves at 15 and 43 GHz\footnote{All the values at 43 GHz
    presented in this paper are from the Large VLBA Project:
  Total \& Polarized Intensity Images of Gamma-Ray Bright Blazars at 43
  GHz, and available at
  http://www.bu.edu/blazars/VLBA$_{-}$GLAST/1510.html.} have been
compared. From Fig. \ref{peak_flux} we see that the flux densities at both
frequencies have a similar behaviour, where the 43-GHz data points
seem to anticipate those at lower frequency. 
Interestingly, just after
March 2009, when several $\gamma$-ray flares have been detected
\citep{dammando11}, the flux density at 43 GHz becomes higher than
that at 15 GHz implying a change in the opacity of the component. The
maximum of the radio emission 
is then reached in April 2009 when another strong $\gamma$-ray
flare was detected \citep{cutini09}. 
In correspondence of this luminosity enhancement
the polarization percentage drops while the EVPA changes by about
75$^{\circ}$(Fig. \ref{multi_chi}) at both 15 and 43 GHz becoming
parallel to the jet direction. In the same period a strong rotation
of the optical polarization vector \citep{marscher10} has been detected
together with the ejection of a new superluminal component observed at
15 GHz (Fig. \ref{fit_mojave}), and already reported by
\citet{abdo10b}, 
and at 43 GHz \citep{marscher10}. A similar example of dramatic
rotation of the polarization angle and a drop of the fractional
polarization in coincidence with a $\gamma$-ray
flare was found in the blazar 3C\,279, indicating a co-spatiality of
the optical and $\gamma$-ray emission region \citep{abdo10c}.
  This has been explained assuming a non-axisymmetric structure
  of the emitting region, likely a curved jet trajectory, 
  rather than a perpendicular shock moving in an
  axially symmetric jet as we suggested in Section 4.1. In this case,
  the degree and angle of polarization strictly depends on the
  instantaneous angle formed by the direction of the motion with our
  line of sight \citep{abdo10c}.
The similarity between 3C\,279 and PKS\,1510-089 suggests a common
origin also for the emission at high energy and in the radio band, 
and the time lag in the flux density behaviour may be
due to opacity effects as the shock passes by. A cospatiality of
  the $\gamma$-ray and radio emitting region was also claimed for the BL
  Lac object OJ\,287, where two $\gamma$-ray flaring episodes 
occurred close in
  time with two major millimeter outbursts \citep{agudo11}. \\
In PKS\,1510-089, the variation in
the polarization angle at both 15 and 43 GHz suggests a change in the
magnetic field orientation in the compact component
rather than Faraday effects caused by an
external screen. 
Changes in the polarization angle of the same
magnitude at different wavelengths were reported by \citet{homan02a} who
monitored the behaviour of a sample of 12 blazars by means of
dual-frequency VLBA observations.\\
It is worth noting that, as in the case of 3C\,279 
where no changes in the radio band
could be found related to the $\gamma$-ray flare \citep{abdo10c}, in
PKS\,1510-089 no 
obvious connection between the $\gamma$-ray activity detected in March
2009 \citep{dammando11} and the radio flux density behaviour has been
found, suggesting that during this flare the synchrotron
  radiation in the radio band is not yet fully optically thin.

\section{Conclusions}
We have presented results from the analysis of multi-epoch
polarimetric 
VLBI, Space-VLBI and archival VLBA data from the MOJAVE programme 
of the flat spectrum radio quasar
PKS\,1510-089 spanning over 15 years (1995-2010). 
This source shows a pc-scale core-jet structure where
superluminal knots are ejected at different times. From the
multi-epoch observations we found that the emission of new blobs 
is roughly constant with a time lag of about one year.
Furthermore, the various jet
components are moving away from the core with an apparent superluminal
speed in the range between 15$c$ and 20$c$ and roughly with the same
position angle, suggesting that the precession of the jet is not
relevant on the timescale of decades in our frame. 
Both the total intensity and the polarized flux density of the core
component show high level of variability. Our analysis shows that
occasionally the
EVPA has abrupt changes of about 90 degrees becoming
roughly perpendicular to the jet direction during episodes related to
enhanced radio emission and opacity variations. These properties may be
explained assuming a change between the optically-thick and
optically-thin regime as a consequence of a shock that varies the opacity.
In this context, the luminosity 
locally increases due to the compression of
the plasma in the direction of the shock propagation, 
causing the amplification of the perpendicular component of the
magnetic field with respect to the parallel one, and the EVPA becomes 
parallel to the direction of the shock propagation. The observed
  properties may also be explained as due to a highly ordered magnetic
  field produced by either an oblique shock or a transverse
one propagating down a jet with a curved
  trajectory. In this case
  the angle of the electric vector 
and level of polarization strictly depend on the instantaneous
angle formed by the direction of the shock with our line of sight,
thus explaining changes in the polarization angle different from the 90
degrees expected during the transition between the opacity regimes.
However, the aforementioned scenarios cannot 
describe the source behaviour after all the detected
flares and more monitoring multiwavelength campaigns are needed to
properly understand the physical processes occurring in this source.\\

\section*{Acknowledgments}

We thank the anonymous referee for reading the manuscript carefully 
and making valuable suggestions. The authors are grateful to
G. Brunetti for fruitful discussion.
This research has made use of data from the MOJAVE database that 
is maintained by the MOJAVE team (Lister et al., 2009, AJ, 137, 3718).
The VLBA and VLA are operated by the 
US National Radio Astronomy Observatory which is a facility of the National
Science Foundation operated under cooperative agreement by Associated
Universities, Inc. This research has made use of the NASA/IPAC
Extragalactic Database NED which is operated by the JPL, Californian
Institute of Technology, under contract with the National Aeronautics
and Space Administration.

\appendix
\section{MOJAVE multi-epoch monitoring (1995-2010)}
\begin{figure}
\begin{center}
\includegraphics{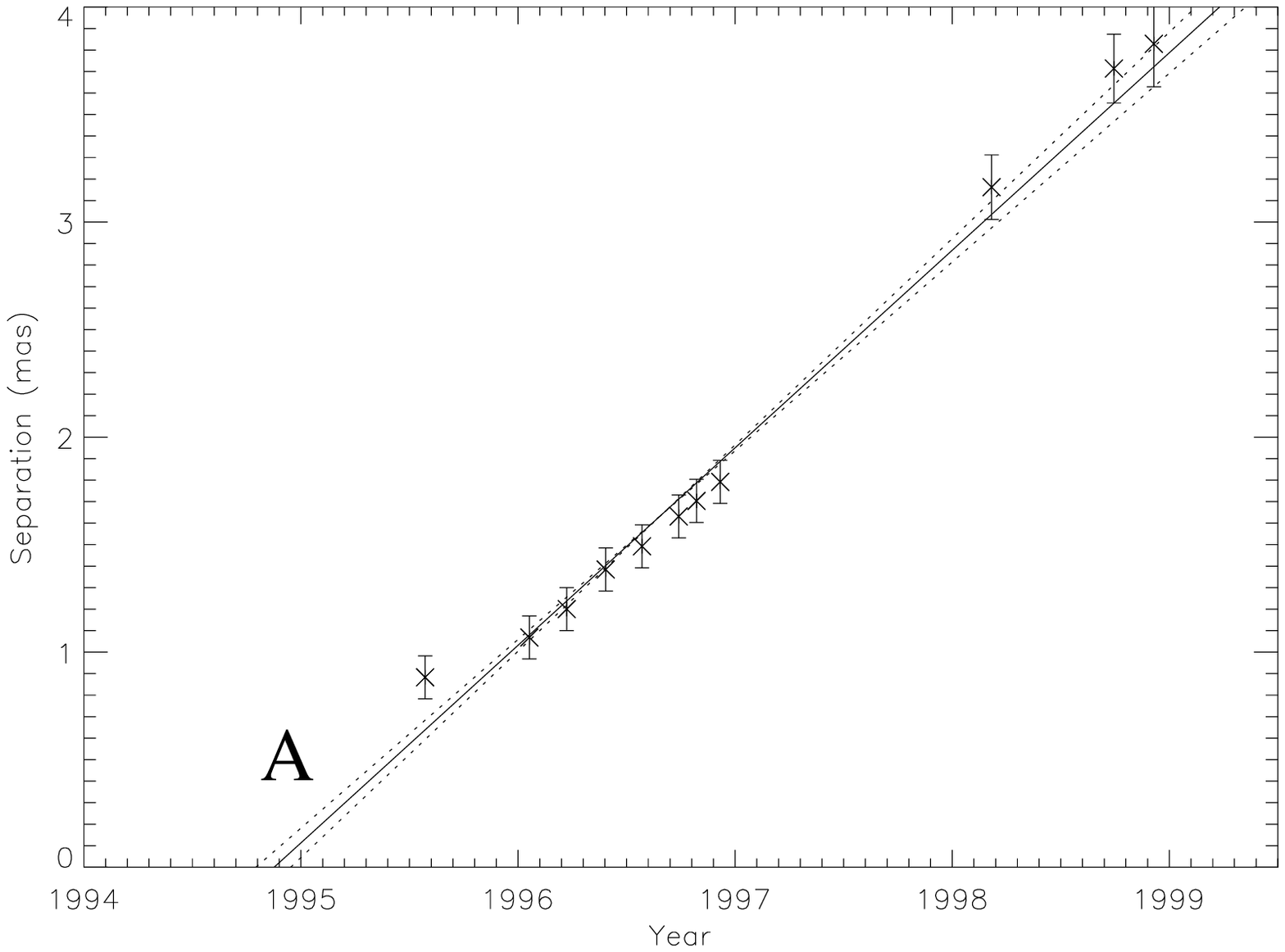}
\vspace{6cm}
\caption{Changes in separation with time between the core and the jet 
component
  A. The solid line represents the regression fit to the data, while
  the dashed-lines represent the uncertainties from the fit parameters. Error
  bars are determined as in Section 3.3.}
\label{fit_1995}
\end{center}
\end{figure}

\begin{figure}
\begin{center}
\includegraphics{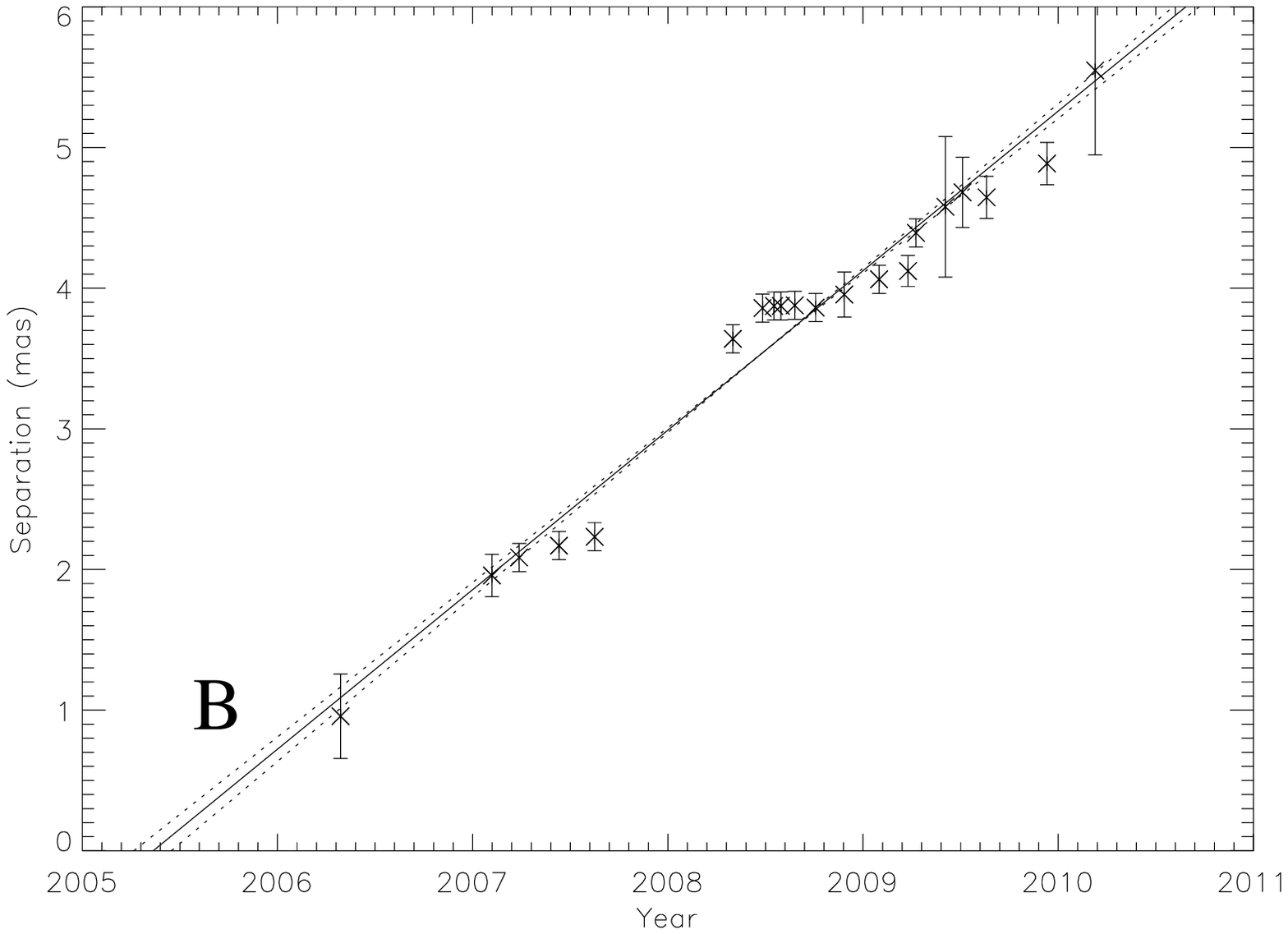}
\vspace{6cm}
\caption{Changes in separation with time between the core and the jet component
  B. The solid line represents the regression fit to the data, while
  the dashed-lines represent the uncertainties from the fit parameters. Error
  bars are determined as in Section 3.3.}
\label{fit_2005}
\end{center}
\end{figure}

To better characterize the evolution of the source structure with a
unique approach we re-analysed VLBA data at 15 GHz from the MOJAVE
programme already published by
\citet{lister09b} and \citet{homan01}. 
Data from each epoch were modelfitted as
described in Section 3.3, providing a homogeneous set of observations
with the same data analysis procedure. \\
The identification and monitoring of the source components
is critical when two subsequent epochs are separated by a long
time interval.
Indeed, if the observation time coverage is sparse, it
becomes very difficult to identify the same component at the various
epochs. By comparing the multi-epoch visibility data 
we found that the source components
could be reliably followed only in the observations from 1995 July to
1998 December, and since 2007.  \\
Between 1995 and 1998 we could identify and follow a jet knot for
which we derive an  
angular separation rate of 0.918$\pm$0.041 mas/yr,
corresponding to an apparent separation velocity $\beta_{\rm app} =
15.0 \pm 0.7$. For this component \citet{homan01} derived an apparent
speed of 15.7$\pm$0.4 if converted in the cosmology used in this paper.
From the linear back-extrapolation fit (Fig. \ref{fit_1995})
we estimate that the
jet component was originated in 1994.92$\pm$0.18. \\  
Between 2007 and 2010, the pc-scale radio morphology displayed a
core-jet structure with the presence of four knots, three of which
have been discussed in Section 3.3. The other knot
we found increases its separation to the core
with a rate of 1.134$\pm$0.035,   
corresponding to an apparent velocity
$\beta_{\rm app}= 18.6\pm0.5$, and it should have been originated 
in 2005.42$\pm$0.11 (Fig. \ref{fit_2005}). For this component
\citet{lister09b} derived a separation speed of 17.0$\pm$2.6, and the
ejection date 2005.87$\pm$0.25.
The kinematic properties of the jet components are reported in Table \ref{tabella_moto}.\\
Between 1999 and 2006 the observations are too sparse in time to allow
a reliable 
identification of the same source component throughout the various data sets.\\

\end{document}